\documentclass[fleqn,twoside]{article}
\usepackage{espcrc2}

\usepackage{graphicx}
\usepackage[figuresright]{rotating}

\newcommand{\AmS}{{\protect\the\textfont2
  A\kern-.1667em\lower.5ex\hbox{M}\kern-.125emS}}
\renewcommand{\theequation}{\arabic{equation}}

\def\beq{\begin{equation}}
\def\eeq{\end{equation}}
\def\bea{\begin{eqnarray}}
\def\eea{\end{eqnarray}}
\def\nn{\nonumber}
\def\dfrac{\displaystyle\frac}   

\def\etal{{\it et al.}}
\def\ibid{{\it ibid.}}
\def\eg{{\it e.g.~}}

\def\PR#1#2#3{Phys. Rev. {\bf #1}, #2 (#3)}
\def\PRL#1#2#3{Phys. Rev. Lett. {\bf #1}, #2 (#3)}
\def\PL#1#2#3{Phys. Lett. {\bf #1}, #2 (#3)}

\def\NP#1#2#3{Nucl. Phys. {\bf #1}, #2 (#3)}

\def\PTP#1#2#3{Prog. Theor. Phys. {\bf #1}, #2 (#3)}
\def\EPJ#1#2#3{Eur. Phys. J. {\bf #1}, #2 (#3)}

\def\PLB#1#2#3{Phys. Lett. {\bf B#1} (#2) #3}

\def\eqref#1{eq.(\ref{eqn:#1})}
\def\eqlab#1{\label{eqn:#1}}

\def\bmaT{\left(\begin{array}{ccc}}
\def\emaT{\end{array}\right)}
\def\bma{\left( \begin{array} }
\def\ema{\end{array} \right)}

\def\ov{\overline}
\def\wt{\widetilde}
\def\l{\left}
\def\r{\right}
\def\gsim{~{\rlap{\lower 3.5pt\hbox{$\mathchar\sim$}}\raise 1pt\hbox{$>$}}\,}
\def\lsim{~{\rlap{\lower 3.5pt\hbox{$\mathchar\sim$}}\raise 1pt\hbox{$<$}}\,}

\makeatletter
\newtoks\@stequation
\def\subequations{\refstepcounter{equation}%
  \edef\@savedequation{\the\c@equation}%
  \@stequation=\expandafter{\theequation}%   %only want \theequation
  \edef\@savedtheequation{\the\@stequation}% %expanded once
  \edef\oldtheequation{\theequation}%
  \setcounter{equation}{0}%
  \def\theequation{\oldtheequation\alph{equation}}}
\def\endsubequations{%
  \ifnum\c@equation < 2 \@warning{Only \the\c@equation\space subequation
    used in equation \@savedequation}\fi
  \setcounter{equation}{\@savedequation}%
  \@stequation=\expandafter{\@savedtheequation}%
  \edef\theequation{\the\@stequation}%
  \global\@ignoretrue}
\def\eqnarray{\stepcounter{equation}\let\@currentlabel\theequation
\global\@eqnswtrue\m@th
\global\@eqcnt\z@\tabskip\@centering\let\\\@eqncr
$$\halign to\displaywidth\bgroup\@eqnsel\hskip\@centering
     $\displaystyle\tabskip\z@{##}$&\global\@eqcnt\@ne
      \hfil$\;{##}\;$\hfil
     &\global\@eqcnt\tw@ $\displaystyle\tabskip\z@{##}$\hfil
   \tabskip\@centering&\llap{##}\tabskip\z@\cr}
\makeatother
\def\bseq{\begin{subequations}}
\def\eseq{\end{subequations}}
\def\cU#1#2{U_{#1}^{#2}}
\def\cUm#1#2{\wt{U}_{#1}^{#2}}
\def\satms#1{\sin^2\theta_{_{\rm ATM}}^{#1}}

\def\srct#1{\sin^22\theta_{_{\rm RCT}}^{#1}}
\def\matm#1{\delta m^{2~#1}_{_{\rm ATM}}}

\def\thatm{\theta^{}_{_{\rm ATM}}} 
\def\thrct{\theta^{}_{_{\rm RCT}}} 
\def\thsol{\theta^{}_{_{\rm SOL}}} 
\def\dmsqatm{\delta m^2_{_{\rm ATM}}} 
\def\dmsqrct{\delta m^2_{_{\rm RCT}}} 
\def\dmsqsol{\delta m^2_{_{\rm SOL}}} 
\def\dmns{\delta_{_{\rm MNS}}^{}}
\def\jmns{J_{_{\rm MNS}}^{}}
\def\cerenkov{$\check{\rm C}$erenkov~}
\def\chisqmin{$\chi^2_{\mbox{\scriptsize \rm min}}$}

\def\cc{\mbox{\scriptsize \rm CC}}
\def\obs{\mbox{\scriptsize \rm obs}}
\def\true{{\mbox{\scriptsize \rm true}}}
\def\mns{{\mbox{\tiny \rm MNS}}}
\def\rct{{\mbox{\tiny \rm RCT}}}
\def\atm{{\mbox{\tiny \rm ATM}}}

\def\diag{\mbox{\rm diag}}

\def\ev{\mbox{\rm eV}}
\title{Physics prospects of future neutrino oscillation experiments in Asia}

\author{Kaoru Hagiwara \address{Theory Division, KEK, Tsukuba 305-0801 Japan}} 
       
\begin{document}

\begin{abstract}
The three neutrino model has 9 physical parameters, 3 neutrino masses, 
3 mixing angles and 3 CP violating phases.  Among them, neutrino 
oscillation experiments can probe 6 parameters: 
2 mass squared differences, 3 mixing angles, and 1 CP phase.  
The experiments performed so far determined the magnitudes of the 
two mass squared differences, the sign of the smaller mass squared 
difference, the magnitudes of two of the three mixing angles, and 
the upper bound on the third mixing angle. 
The sign of the larger mass squared difference (the neutrino mass 
hierarchy pattern), the magnitude of the third mixing angle and 
the CP violating phase, and a two-fold ambiguity in the mixing 
angle that dictates the atmospheric neutrino oscillation  
should be determined by future oscillation experiments.  
In this talk, I introduce a few ideas of future long baseline 
neutrino oscillation experiments which make use of the super 
neutrino beams from J-PARC (Japan Proton Accelerator Research 
Complex) in Tokai village.  
We examine the potential of HyperKamiokande (HK), the proposed 
1 Mega-ton water \cerenkov detector, and then study the fate and 
possible detection of the off-axis beam from J-PARC in Korea, 
which is available free throughout the period of the T2K 
(Tokai-to-SuperKamiokande) and the possible T-to-HK projects.  
Although the CP violating phase can be measured accurately by studying 
$\nu_\mu \to \nu_e$ and $\ov\nu_\mu \to \ov\nu_e$ oscillations 
at HK, there appear multiple solution ambiguities which can be 
solved only by determining the neutrino mass hierarchy and the 
twofold ambiguity in the mixing angle.  
We show that very long baseline experiments with higher energy beams 
from J-PARC and a possible huge Water \cerenkov Calorimeter detector 
proposed in Beijing can resolve the neutrino mass hierarchy.  
If such a detector can be built in China, future experiments 
with a muon storage ring neutrino factory at J-PARC will be able to 
lift all the degeneracies in the three neutrino model parameters.  
\vspace{1pc}
\end{abstract}

% typeset front matter (including abstract)
\maketitle

\section{Introduction}

All the experimental observations on neutrino physics, with a 
notable exception of the LSND experiment \cite{lsnd}, can be 
accommodated well in the three neutrino model.  
The atmospheric neutrino oscillation first observed by 
Super-Kamiokande in 1998 \cite{sk98} and the observed deficit of 
the muon neutrino flux in the K2K experiment \cite{k2k03} can be interpreted 
as $\nu_\mu \to \nu_\tau$ oscillation with the 90\% CL allowed ranges 
\cite{sk-atm04}
\bea
\label{data-atm}
1.9 \times 10^{-3} < &\dmsqatm (\mbox{\rm eV}^2)& < 3.0 \times 10^{-3} 
\,, \nn\\
&\sin^2 2\thatm& > 0.90 \,.
\eea
The observed deficits of the solar neutrinos \cite{kayser03} and 
the reactor anti-neutrinos by the KamLand experiment can be interpreted 
as $\nu_e \to \nu_\mu$ or $\nu_\tau$ oscillation with \cite{kamland04} 
\bea
\label{data-sol}
\dmsqsol &=& 8.2^{+0.6}_{-0.5} \times 10^{-5} \mbox{\rm eV}^2 \,, \nn\\
\tan^2 \thsol &=& 0.40^{+0.09}_{-0.07} \,.
\eea
No observation of the deficit of the reactor $\ov\nu_e$ 
flux constrains the third mixing angle \cite{chooz}
\bea
\label{data-rct}
\sin^2 2\thrct < 0.20 \:\: \mbox{\rm for} \:\: 
\delta m^2 = 2.0 \times 10^{-3} \mbox{\rm eV}^2 \,,\nn\\ 
\sin^2 2\thrct < 0.16 \:\: \mbox{\rm for} \:\: 
\delta m^2 = 2.5 \times 10^{-3} \mbox{\rm eV}^2 \,, \\ 
\sin^2 2\thrct < 0.14 \:\: \mbox{\rm for} \:\: 
\delta m^2 = 3.0 \times 10^{-3} \mbox{\rm eV}^2 \,,\nn
\eea
at the 90\% CL.  
Since the magnitude of the mass squared difference of order 1~eV$^2$ 
associated with the $\nu_\mu \to \nu_e$ oscillation observed by 
the LSND experiment is much bigger than those of the above differences, 
once their finding is confirmed by the MiniBoone experiment \cite{miniboone}, 
we need at least one more neutrinos.  
Nevertheless, I would like to discuss physics prospects of future 
neutrino oscillation experiments in the framework of the three neutrino 
model, mainly because of my prejudice that I find no compelling theoretical 
reason for the existence of more than three light neutrinos.  
Therefore, once the LSND observation is confirmed by MiniBoone, 
we should revise all our thoughts from scratch.  

In the three neutrino model, the three weak interaction eigenstates 
$\nu_\alpha$ ($\alpha = e, \mu, \tau$) that appear in the universal 
charged current interactions and the three mass eigenstates 
$\nu_i$ ($i=1,2,3$) are related by the $3\times3$ MNS 
(Maki-Nakagawa-Sakata) matrix \cite{mns} 
\bea
\label{numixing}		
\nu_\alpha^{}=
\sum_{i=1}^3 \l({{V}_{_{\rm MNS}}}\r)_{{\alpha}{i}}^{}
~{\nu_i^{}}\,.
\eea
The MNS matrix can be expressed as \cite{kayser03}
\bea
\label{v-mns}
V_{_{\rm MNS}}^{} 
= U {\cal P} 
= U \mbox{diag}(e^{i \alpha_1^{}/2},e^{i \alpha_2^{}/2},1)
\eea
with the two Majorana phases $\alpha_1$ and $\alpha_2$.  
The 3$\times$3 unitary matrix $U$ can be parameterized in terms of 
the three angles and one CP violating phase just like the CKM matrix:
\bea
\label{u-mns}
U &=& O_{23} P_\delta^{} O_{13} P_\delta^\dagger O_{12} \nn\\
&=& 
% \left(\begin{array}{ccc}
%   c_{12}c_{13}     & s_{12}c_{13}    & s_{13}e^{-i\delta} \\
%  -s_{12}c_{23}-c_{12}s_{23}s_{13}e^{i\delta} & 
% ~~c_{12}c_{23}-s_{12}s_{23}s_{13}e^{i\delta} & s_{23}c_{13} \\
% ~~s_{12}s_{23}-c_{12}c_{23}s_{13}e^{i\delta} & 
%  -c_{12}s_{23}-s_{12}c_{23}s_{13}e^{i\delta} & c_{23}c_{13}  
% \end{array}\right) \nn \\
% &=&
\left(\begin{array}{ccc}
U_{e 1}    & U_{e 2}    & U_{e 3} \\
U_{\mu 1}  & U_{\mu 2}  & U_{\mu 3} \\
U_{\tau 1} & U_{\tau 2} & U_{\tau 3}  
\end{array}\right) \,,
\eea
where $O_{ij}$ are the orthogonal rotation matrix in the $ij$ plane, 
and $P_\delta = \mbox{\rm diag}(1,1,e^{i\delta})$ gives the CP 
violating phase of the MNS matrix $\delta = \delta^{}_{_{\rm MNS}}$.  
Because the present neutrino oscillation experiments
constrain directly the three elements,
\bseq
\label{ho98-vs-pdg} 
\bea
U_{e3}    &=& s_{13}e^{-i\delta} \,,\\
U_{e2}    &=& s_{12}c_{13}       \,,\\
U_{\mu 3} &=& s_{23}c_{13}       \,,
\eea
\eseq
we find it most convenient to adopt the convention \cite{ho98}
where these three matrix elements in the upper-right
corner of the $U$ matrix are chosen as the independent parameters.
The phase convention of Refs.~\cite{ho98,t2b}
\bseq
\label{ho98convention}
\bea
&& U_{e3} = |U_{e3}|e^{-i\delta} \,,\\
&& U_{e1} \geq U_{e2}    \ge 0 \,,\\
&& U_{\mu 3}, U_{\tau 3} \ge 0 \,,
\eea
\eseq
is equivalent to the choice $0\le \theta_{12} \le \pi/4$ and 
$0\le \theta_{13},\theta_{23} \le \pi/2$.
In our convention $\nu_1$ is defined by the condition $U_{e1}>U_{e2}$, 
and hence both $m_1 < m_2$ and $m_1 > m_2$ are allowed, and they 
should be determined by experiments.\footnote{
This is equivalent to an alternative definition of $\nu_1$ where 
$m_1 < m_2$ is assumed and the ordering $U_{e1}>U_{e2}$ or 
$U_{e1}<U_{e2}$ is to be determined by experiments.  
}%
Four possible patterns of the neutrino mass hierarchy are then 
defined as \cite{t2b}
\bea
\label{hierarchy}
&& (\delta m^2_{12},\delta m^2_{13}) \\
&& = (\hphantom{-}\dmsqsol,\hphantom{-}\dmsqatm) 
\:\:\: \mbox{\rm (I: normal hierarchy)} \nn \\
&& = (-\dmsqsol,\hphantom{-}\dmsqatm) 
\:\:\: \mbox{\rm (II)}  \nn\\
&& = (\hphantom{-}\dmsqsol,-\dmsqatm) 
\:\:\: \mbox{\rm (III: inverted hierarchy)} \nn \\
&& = (-\dmsqsol,-\dmsqatm) 
\:\:\: \mbox{\rm (IV)} \nn 
\eea
where $\delta m^2_{ij} = m_j^2 - m_i^2$.  
See Fig.~1.  
The solar neutrino experiments determine 
\bea
\label{dmsqsol}
m^2_{12} \equiv m_2^2 - m_1^2 = \dmsqsol > 0 \,, 
\eea
and hence only the hierarchy patterns I (normal) or III (inverted) 
are yet to be determined by future experiments.  

\begin{figure}[t]%[thbp]
%\begin{center}
\includegraphics[angle=-90,width=8cm]{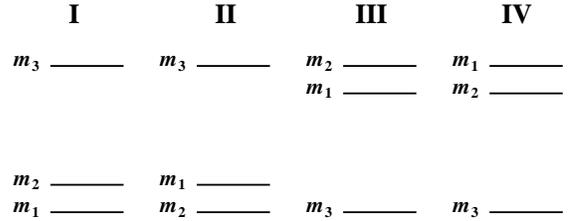}
%\end{center} 
\caption{Schematic view of the four-types of neutrino-mass hierarchy.
}
\label{fig:hi-1234}
\end{figure}

We keep the four hierarchy cases in the following discussions because 
we find the following theorem \cite{t2b,kh02} useful: 
\bseq
\label{hi-theorem}
\bea
P_{\ov\nu_\alpha\to\ov\nu_\beta}^{I}\: = 
P_{\nu_\alpha\to\nu_\beta}^{IV} &&\;(\mbox{\rm normal hierarchy}) 
\,,\quad\\ 
P_{\ov\nu_\alpha\to\ov\nu_\beta}^{III} = 
P_{\nu_\alpha\to\nu_\beta}^{II} &&\;(\mbox{\rm inverted hierarchy}) 
\,.\quad
\eea
\eseq
The theorem is valid in the presence of an arbitrary matter profile 
\cite{kh02}.  
Because of the theorem (\ref{hi-theorem}), we do not give expressions 
for the anti-neutrino oscillation probabilities explicitly.  
For instance, $P_{\ov\nu_\mu \to \ov\nu_e}$ in the normal hierarchy 
is obtained from the expression for $P_{\nu_\mu \to \nu_e}$ simply 
be reversing the sign of the factors $\Delta_{12}$ and $\Delta_{13}$.  

The probability of finding the flavor-eigenstate $\beta$
from the original flavor-eigenstate $\alpha$ at the base-line length $L$
in the vacuum is given by
\bea
\label{pnua2nub}
&&P_{\nu_\alpha \to \nu_\beta} \\
&=&
\left|\sum_{j=1}^3 (V_{_{\rm MNS}})_{\beta j}^{} 
\exp\left(-i\frac{m_j^2}{2E_\nu}L\right) (V_{_{\rm MNS}})_{j \alpha} 
\right|^2\nn \\
\nn \\
&=&
\left| \cU{\beta 1}{} 
 \cU{\alpha 1}{\ast}
+\cU{\beta 2}{} e^{-i\Delta_{12}} \cU{\alpha 2}{\ast} 
+\cU{\beta 3}{} e^{-i\Delta_{13}} \cU{\alpha 3}{\ast}
\right|^2 \,, 
\nn
\eea
where $\Delta_{ij}$ is
\bea
\label{delta-ij}
\Delta_{ij} \equiv \frac{\delta m_{ij}}{2E_\nu}L
\simeq 2.534 \frac{\delta m_{ij}^2 ({\rm eV}^2)}{E_\nu({\rm GeV})}L({\rm km}
)\,.  
\eea 
It is clear from this expression that the neutrino oscillation experiments 
cannot measure the two Majorana phases $\alpha_1$ and $\alpha_2$ in 
eq.~(\ref{v-mns}), and that only the two mass squared differences out 
of the three neutrino masses can be probed by those experiments.  

It is useful to give expressions of the oscillation probabilities 
when either $\Delta_{13}$ or $\Delta_{12}$ is order unity.
For instance when $|\Delta_{13}| \sim O(1)$, we can expand the 
expressions in terms of $|\Delta_{12}| \ll 1$ and find 
\bseq
\label{pnua2nub-2}
\bea
&&
P_{\nu_\mu \to \nu_\mu} 
= 1 -4|U_{\mu 3}|^2(1-|U_{\mu 3}|^2) \sin^2\frac{\Delta_{13}}{2} 
\quad\quad\nn\\
&&
+ 2|U_{\mu 2}|^2|U_{\mu 3}|^2 \sin\Delta_{13}\cdot \Delta_{12} 
+O((\Delta_{12})^2) \,,
\label{pnumu2numu}
\\
&&
P_{\nu_e \to \nu_e}
= 1 -4|U_{e3}|^2(1-|U_{e3}|^2) \sin^2\frac{\Delta_{13}}{2} 
\nn\\
&&
+ 2|U_{e2}|^2|U_{e3}|^2 \sin\Delta_{13}\cdot \Delta_{12} 
+O((\Delta_{12})^2) \,,
\label{pnue2nue}
\\
&&
P_{\nu_\mu \to \nu_e}
= 4|U_{\mu 3}|^2|U_{e3}|^2\sin^2\frac{\Delta_{13}}{2} 
\nn\\
&&
+2Re[U_{\mu 2}U_{e2}^*U_{e3}U_{\mu 3}^*]\sin\Delta_{13}\cdot\Delta_{12} 
\nn\\
&&
+4\jmns\sin^2\frac{\Delta_{13}}{2}\cdot\Delta_{12}
+O((\Delta_{12})^2) \,.
\label{pnumu2nue}
\eea
\eseq
The expression eq.~(\ref{pnumu2numu}) is relevant for the atmospheric 
neutrino oscillation and the K2K experiment, and we may identify 
\bseq
\label{atm}
\bea
\sin^2 2\thatm &=& 4 |U_{\mu 3}|^2 ( 1 - |U_{\mu 3}|^2) \,,
\\
\dmsqatm &=& |\delta m^2_{13}| = |m_3^2-m_1^2| \,.
\eea
\eseq
The expression eq.~(\ref{pnue2nue}) is relevant for the reactor 
anti-neutrino oscillation experiments after the replacements in 
eq.~(\ref{hi-theorem}) is made:
\bseq
\label{rct}
\bea
\sin^2 2\thrct &=& 4 |U_{e3}|^2 ( 1 - |U_{e3}|^2) \,,
\\
\dmsqrct &=& |\delta m^2_{13}| = \dmsqatm \,.
\eea
\eseq
Only the magnitude of $\delta m^2_{13}$ is constrained because, 
so far no effects of order $\Delta_{12}$ have been positively 
identified.  
Finally, eq.~(\ref{pnumu2nue}) is relevant for the $\nu_\mu \to \nu_e$ 
appearance experiments such as T2K.  It is important to note that 
the $\nu_\mu \to \nu_e$ oscillation probability is proportional to 
\bea
\label{numu2nuefactor} 
4~\sin^2\thatm~\sin^2\thrct = 4|U_{\mu 3}|^2|U_{e3}|^2 \,.
\eea
and that the CP violating effect appears in the $\Delta_{12}$ order.        
The Jarlskog parameter of the MNS matrix is defined as
\bseq
\label{jmns}
\bea
\jmns &\equiv& 
{\rm Im} \left({U_{\alpha i} U_{\beta i}^{\ast} U_{\beta j} 
U_{\alpha j}^{\ast}}\right) 
\\
&=& - \frac{U_{e1}U_{e2}U_{\mu 3}U_{\tau 3}}{1-\left|U_{e3}\right|^2}
              {\rm Im}\left({U_{e3}}\right)
\label{jmns-ho98}
\\
&=& c_{13}^2 c_{12} c_{23} s_{12} s_{23} s_{13} \sin\dmns 
\label{jmns-para}
\eea
\eseq
where the orderings $(\alpha, \beta)=(e, \mu),(\mu, \tau),(\tau, e)$ 
and $(i,j)=(1,2),(2,3),(3,1)$ should be taken in the definition.  
The expression (\ref{jmns-ho98}) follows from the convention 
(\ref{ho98convention}) \cite{ho98}, and eq.~(\ref{jmns-para}) follows 
from the parameterization (\ref{u-mns})~\cite{kayser03}.     

Comparison of the expansions in eq.~(\ref{pnua2nub}) and the 
theorem (\ref{hi-theorem}) tells that the anti-neutrino oscillation 
probabilities have the same expressions as those for neutrinos 
up to terms of oder $\Delta_{12}$, except for the term proportional 
to $\jmns$.  We find 
\bea
\label{diff-cpv}
&&P_{\nu_\mu \to \nu_e} - P_{\ov\nu_\mu \to \ov\nu_e} 
\nn\\
=&& 8\jmns \sin^2\frac{\Delta_{13}}{2}\cdot\Delta_{12} 
+O((\Delta_{12})^2) \,,
\quad
\eea
which is valid for both normal and inverted hierarchies.  
Although the above expression is valid only when we can neglect the 
earth matter effects of the oscillation probabilities, 
it gives a good zeroth order approximation to the proposed T2K and 
T-to-HK experiments, which plan to use sub-GeV neutrino beams.    

When the magnitude of $\Delta_{12}$ is of the order of unity, 
$|\Delta_{13}| \sim \dmsqatm/\dmsqsol \sim 30 \gg 1$ applies, 
and we have an alternative expansion of eq.~(\ref{pnua2nub}).  
The most relevant one is  
\bea
\label{pnue2nue2}
P_{\nu_e \to \nu_e} &=& 1 
-4|U_{e1}|^2|U_{e2}|^2 \sin^2\frac{\Delta_{12}}{2} 
\qquad\qquad\nn\\ 
&& -\frac{1}{2}\sin^2 2\thrct 
+O((\frac{E_\nu}{\delta E_\nu \Delta_{13}}))\,, 
\eea
where $\delta E_\nu/E_\nu$ is the experimental resolution of the 
neutrino energy.  
The expression (\ref{pnue2nue2}) is relevant for the solar neutrino 
oscillation and the KamLand reactor neutrino oscillation experiments.  
By neglecting the small term proportional to $\sin^2 2\thrct$, 
we can identify 
\bseq
\label{sol} 
\bea
\sin^2 2\thsol &=& 4|U_{e2}|^2|U_{e1}|^2 \,, 
\\
\dmsqsol &=& |\delta m^2_{12}| = |m_2^2 - m_1^2| \,.
\eea
\eseq  
The sign of the mass squared difference cannot be determined from 
the above expression, which applies for the KamLand experiments.  

From the above arguments, we can parameterize the complete 
MNS matrix elements by using the observed constraints on the 
three mixing angles:
\bea
\label{3angles}
|U_{e3}|^2 &=& \frac{1-\sqrt{1-\sin^2 2\thrct}}{2} \,,
\nn\\
U_{e2}^2 &=& 
\frac{1-|U_{e3}|^2-\sqrt{(1-|U_{e3}|^2)^2-\sin^22\thsol}}{2} \,,
\nn\\
U_{\mu 3}^2 &=& \frac{1\pm\sqrt{1-\sin^2 2\thatm}}{2} \,.
\eea
Since the magnitudes of the three matrix elements at the corner 
of the MNS matrix are determined as above, the full MNS matrix 
can be determined for a given CP phase $\dmns$ \cite{ho98}.  
Although the above identifications are valid only approximately, 
they suffice for our discussions about physics prospects of 
future neutrino oscillation experiments.  

\section{Matter effects in the neutrino oscillation} 

The neutrino mass hierarchy pattern between I and II (III and IV) has 
been distinguished by the matter effects of the observed solar 
neutrino oscillation.  

The oscillation probabilities in the matter has the same form as 
in eqs.~(\ref{pnua2nub}), (\ref{delta-ij}) and (\ref{pnua2nub-2}), 
where the mass-squared differences and the MNS matrix elements are 
replaced by those in the matter:
\bea
\label{replacement}
\Delta_{ij} \to \wt \Delta_{ij}
\,,~~U_{\alpha i} \to \wt U_{\alpha i}
\,,~~J_{_{\rm MNS}} \to \wt J_{_{\rm MNS}}
\,,\quad
\eea
if the variation of the matter density remains small along the base-line.

Neutrino-flavor oscillation inside of the matter is governed by the 
equation \cite{matter,msw}
\bseq
\label{eom-matter}
\bea
\label{eom-matter1}
i\frac{\partial}{\partial t}
\bma{c}
\nu_e \\
\nu_\mu \\
\nu_\tau\\
\ema
=\frac{1}{2E_\nu}
H 
\bma{c}
\nu_e \\
\nu_\mu \\
\nu_\tau\\
\ema
\,,
\eea
where the Hamiltonian (in units of $2E_\nu$) in the matter is 
\bea
\label{h-matter}
H&=&U~\diag(0,\delta m^2_{12}, \delta m^2_{13})~U^\dagger 
\nn\\
&&+\diag(a,0,0) 
\\
&=& 
{{\cUm{}{}}}
\bmaT
 {\lambda_1} & 0 & 0 \\
 0 & {\lambda_2} & 0 \\
 0 & 0 & {\lambda_3}
\emaT
{{\cUm{}{\dagger}}}
\,.
\label{mns-matter}
\eea
\eseq 
Here $a$ is the matter effect term,
\bea
\label{a-matter}
a&=&2\sqrt{2}G_F n_e E_\nu^{} 
\nn\\
&=&
{7.56}\times 10^{-5}({\rm eV}^2)\left(\frac{\rho}{{\rm g/cm}^{3}}\right)
\left(\frac{E_\nu}{\rm GeV}\right)\,,\quad
\eea
where 
$n_e$ is the electron density, 
$G_F$ is the Fermi constant,
and $\rho$ is the matter density.

\begin{figure}[ht]
\includegraphics[width=7cm]{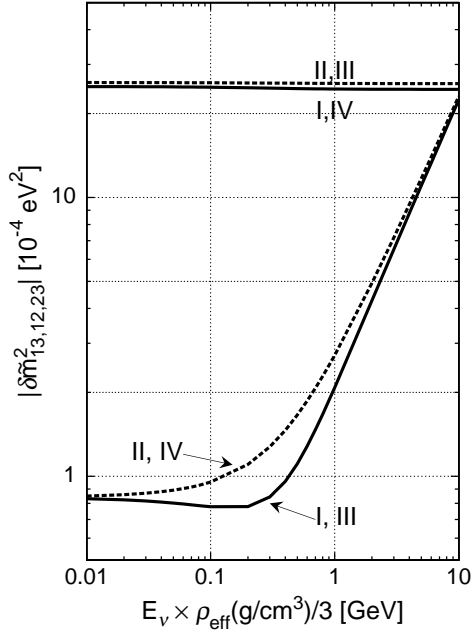}
\caption{%
The magnitudes of the effective mass squared differences, 
$|\delta \wt{m}^2_{13}|$ 
($|\delta \wt{m}^2_{23}|$ in the hierarchy II, IV cases) 
and $|\delta \wt{m}^2_{12}|$,  
as functions of the neutrino energy times the effective matter 
density $\rho_{\rm eff}$(g/cm$^3$)/3.  
In the earth crust, $\rho_{\rm eff}\sim 3$, 
and at the center of the sun, $\rho_{\rm eff}\sim 200$.  
}
\label{fig:matter}
\end{figure}

The Hamiltonian $H$ in the matter is diagonalized 
by the MNS matrix in the matter $\wt U$.
The neutrino-flavor oscillation probabilities in the matter 
take the same form as those in the vacuum by making 
the replacement (\ref{replacement}) where 
\bea
\label{dmsq-m}
\wt{\Delta}_{ij} = 
\dfrac{{\lambda_j}-{\lambda_i}}{2E_\nu}L
\equiv
\dfrac{\delta \wt{m}_{ij}^{2}}{2E_\nu}L\,.
\eea
In Fig.~\ref{fig:matter}, we show the $E_\nu$-dependence 
of the magnitudes of the effective mass-squared differences,  
$|\delta \wt{m}_{13}^{2}|$ and $|\delta \wt{m}_{12}^{2}|$ 
for the 4 mass hierarchy cases of Fig.~\ref{fig:hi-1234}.  
In case of the hierarchy II and IV, we show $|\delta \wt{m}_{23}^{2}|$ 
rather than $|\delta \wt{m}_{13}^{2}|$, because the leading 
oscillation term is dictated by the former difference.   
The curves are obtained for
\bseq
\label{para-set0}
\bea 
&&|\delta m^2_{13}| = 2.5\times 10^{-3}\ev^2 \,,
\\
&&|\delta m^2_{12}| = 8.3\times 10^{-5}\ev^2 \,,
\\
&&\tan^2\thsol = 0.4  \,,
\\
&&\sin^2\thrct = 0 \,.  
\eea
\eseq
It is worth noting that the matter effects depend only on the 
above four parameters, i.e., the two mass squared differences 
and the two angles, $\theta_{12}$ and $\theta_{13}$, because 
of the identity \cite{ykt02}
\bseq
\label{eq:ykt02}
\bea
H = O_{23}P_\delta H' P_\delta^\dagger O_{23}^T \,,
\eea
where 
\bea
H'&=& O_{13}O_{12} \,\diag(0,\delta m^2_{12},\delta m^2_{13})\, 
O_{12}^T O_{13}^T 
\nn\\
&&+ \,\diag(a,0,0) \,. 
\eea
\eseq
It is clear that the above reduced Hamiltonian $H'$ can be 
diagonalized as 
\bseq
\label{eq:ykt02b}
\bea
H' = \wt{O} \,\diag(\lambda_1,\lambda_2,\lambda_3)\, \wt{O}^T \,, 
\eea
with the same eigen values of $H$ in eq.~(\ref{h-matter}).  
The MNS matrix in the matter is then expressed as\footnote{
Note that once the expression (\ref{umns-matter}) is used to 
calculate the MNS matrix in the matter, the phase convention 
of eq.~(\ref{ho98convention}) is no more valid.}  
\bea
\label{umns-matter}
\wt{U} = O_{23} P_\delta \wt{O} \,.
\eea
\eseq
In the limit of $\sin^2\thrct=(s_{13})^2=0$ in eq.~(\ref{para-set0}), 
the matrix $O_{13}=1$, and the eigen value problem is reduced to 
the $2\times 2$ matrix problem.  The eigenvalues are 
\bea
\lambda_\pm &=&  
\frac{\delta m^2_{12}+a\pm\sqrt{
(\delta m^2_{12}+a)^2-4ac_{12}^2\delta m_{12}^2}}{2} \,,
\nn\\
\lambda_3 &=&  \delta m^2_{13} \,,
\label{eq:2by2}
\eea
and the results are shown in Fig.~\ref{fig:matter} for all the 
four hierarchy cases of Fig.~\ref{fig:hi-1234}.  
Note that 
\bseq
\label{lambda12}
\bea
(\lambda_1,\lambda_2) &=& (\lambda_-,\lambda_+) 
\;\mbox{\rm hierarchy I, III} \,, 
\\
&=& (\lambda_+,\lambda_-) 
\;\mbox{\rm hierarchy II, VI} \,,
\eea
\eseq
so that $\lambda_1 \to 0$ and 
$\lambda_2-\lambda_1 = \delta\wt{m}^2_{12} \to \delta m^2_{12}$ 
in the $a \to 0$ limit.  
Although all the curves in Fig.~\ref{fig:matter} are obtained 
in the limit of $\sin^2 \thrct =0$, we find that the qualitative 
features of the curves remain the same even when 
$\sin^2 2\thrct \approx 0.1$, except near the 10~GeV region.   

The horizontal scale of Fig.~\ref{fig:matter} is the neutrino energy 
times the effective matter density in units of 3~g/(cm)$^3$, 
and the vertical scale gives the magnitudes of the effective 
mass squared differences in the matter.  
The typical energy of the reactor anti-neutrinos is 
about a few MeV, or a few $\times 10^{-3}$~GeV, and hence there 
is no significant difference between the hierarchy I and II 
(or between III and IV) in the CHOOZ or KamLand experiments.  
On the other hand, the typical Boron (${}^8$B) neutrino flux 
energy of 7~MeV with the effective matter density of 
$\rho^{}_{\rm eff} < 200$~g/(cm)$^3$ in the sun 
extends up to 450~MeV in 
the horizontal scale, where the matter effects clearly distinguish 
between the neutrino mass hierarchies, and the energy dependence 
of the observed solar neutrino flux clearly favors the hierarchy  
I or III, over II or IV.  
In fact, in case of the hierarchy I or III, the effective 
mass squared difference 
\bea
\label{dmsq12-matter}
&&|\delta \wt{m}^2_{12}| = |\lambda_2 - \lambda_1| 
\\
&&=\sqrt{(a-\cos 2\theta_{12}\,\delta m^2_{12})^2
+(\cos 2\theta_{12}\,\delta m^2_{12})^2}
\nn
\eea
has the minimum at 
\bea
\label{dmsq12-min}
a=\cos 2\theta_{12}\; \delta m^2_{12}
\eea
where the effective mixing angle $\wt{\theta}_{12}$ crosses $\pi/4$.  
On the other hand, in case of the hierarchy II or IV, since 
$\delta m^2_{12}<0$ in eq.~(\ref{dmsq12-matter}), there is no minimum, 
and the effective mixing angle decreases at high energy or density.  
The difference is significant enough to affect the observed 
energy spectrum of the solar neutrino flux on the earth, and the 
neutrino mass hierarchy I or III has been chosen against II or IV. 

Between the hierarchy I (normal) and III (inverted), 
the difference in the matter effects remains small for the 
effective mass squared differences in Fig.~\ref{fig:matter}.  
The ongoing K2K experiment with the baseline length of 250~km 
and the planned long baseline neutrino oscillation experiments T2K,  
from J-PARC at Tokai to Kamioka, whose baseline length is about 300~km, 
adopt rather low energy neutrino beam below 1~GeV, where 
the matter effect is rather small.  
We will find below that the remaining ambiguity between the neutrino 
mass hierarchy I (normal) and III (inverted) can be distinguished 
by using the earth matter effects at higher energy very long 
baseline neutrino oscillation experiments.   

\section{T2K and T-to-HK projects}

The proposed long-baseline (LBL) neutrino-oscillation 
experiments \cite{minos,t2k} will achieve
the precision measurements of $\delta m^2_{_{\rm ATM}}$ and 
$\sin^22\theta_{_{\rm ATM}}$ 
by using conventional neutrino beams,
which are made from decays of $\pi$ and $K$ that are produced 
by high-energy proton beams.  
In particular, the T2K experiment will be able to detect 
$\nu_\mu \to \nu_e$ appearance, if the magnitude of the 
oscillation eq.~(\ref{numu2nuefactor}) is not too small.  

\begin{figure}[ht]
\includegraphics[width=7.5cm]{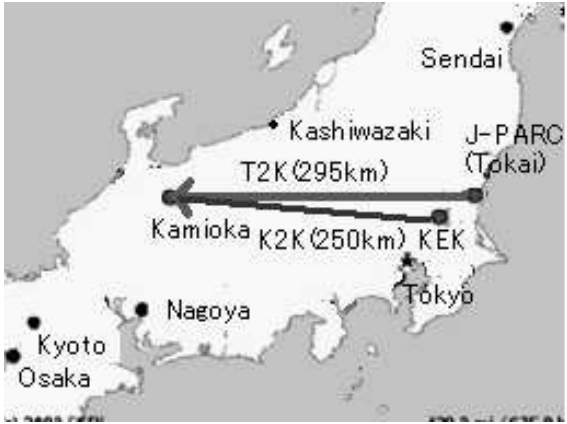}
\caption{%
The long baseline neutrino oscillation projects, 
K2K (KEK to Super-Kamiokande; 250km) and 
T2K (J-PARC at Tokai to Super-Kamiokande; 295km). 
}
\label{fig:k2k-t2k}
\end{figure}

The T2K neutrino beam line is now under construction at J-PARC 
(Japan Proton Accelerator Research Complex) in Tokai village~\cite{jparc},   
whose 50~GeV high intensity proton accelerator with 0.75~MW will deliver 
$10^{21}$ POT (proton on target) per year.  
This is about a factor of 50 more intensive than the KEK proton 
synchrotron for the present K2K experiment.  
Since Tokai village is about 50~km east from KEK, the baseline length  
of the T2K experiment will be $L=295$~km, as compared to $L=250$~km 
for K2K: see Fig.~\ref{fig:k2k-t2k}.  
They plan to make use of the K2K proven technique to reconstruct 
neutrino energy at Super-Kamiokande (SK) by selecting quasi-elastic 
charged current events.  The expected accuracy is 
\bseq
\label{t2k-goal}
\bea
&&\Delta(|\delta m^2_{13}|) = 10^{-4}~\mbox{\rm eV}^2 \,,
\\
&&\Delta(\sin^2 2\thatm) = 0.01 \,,
\eea
\eseq
from the precision measurement of the deficit of the $\nu_\mu$ flux.  
Because SK is capable of detecting $\nu_e$ charged current events 
clearly, they will have a good sensitivity to the $\nu_\mu \to \nu_e$ 
appearance measurement if 
\bea
\label{t2k-numu2nue}
\sin^2 2\thrct \gsim 0.006 \,.
\eea

\begin{figure}[ht]
\begin{center}
\includegraphics[angle=-90,width=7cm]{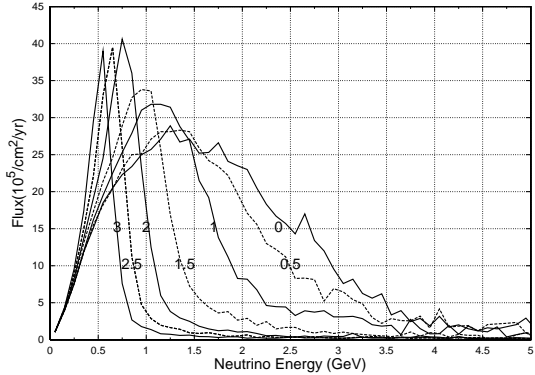}
\end{center}
\caption{%
The neutrino beam flux from J-PARC at L=295km for 
different off-axis angles, 3, 2.5, 2, 1.5, 1, 0.5 and 0 degree 
(on axis).  
}
\label{fig:oab}
\end{figure}

Even with the intensive proton beam from J-PARC, it is not an easy 
task to measure the oscillation probabilities for anti-neutrinos 
because the charged current cross section of anti-neutrinos is 
a factor of 3 smaller than the neutrino cross sections.  
A factor of 5 upgrade in the power of the proton beam to 4MW, 
and a factor of 50 larger detector, 1~Mton water \cerenkov detector 
Hyper-Kamiokande (HK) are therefore envisaged~\cite{HK}.  

In Ref.\cite{t2hk}, we studied the possibility of measuring $\dmns$
in the future T-to-HK experiment.  
There the measurement of the difference (\ref{diff-cpv}) by 
making use of both the neutrino ($\nu_\mu$ enriched) beam and 
the anti-neutrino ($\ov\nu_\mu$ enriched) beam allows us to 
study CP violation in the lepton sector.  
We find that the advantage of the T-to-HK experiment over various 
proposed neutrino oscillation experiments lies in the fact that 
the water \cerenkov detector is capable of measuring both $\mu$ 
and $e$ charged current reactions, and that the use of low energy 
beam (of order 1~GeV or less) at relatively short distances 
(295~km) makes the matter effects small.  
Because of the above advantages, the experiment can effectively 
measure the difference (\ref{diff-cpv}), and hence have 
high sensitivity to the CP violating leptonic Jarlskog parameter
$\jmns$ of eq.~(\ref{jmns}), or $\sin\dmns$.  
For instance, it is relatively easy to distinguish between 
$\dmns=90^\circ$ and $\dmns=270^\circ$.  
On the other hand, we found that it is a non-trivial task for 
the experiment to distinguish between 
\bea 
\label{dmns-pair}
\dmns \quad \mbox{\rm vs} \quad \pi-\dmns \,,
\eea
most notably, between $\dmns=0^\circ$ and $\dmns=180^\circ$.  
In order to distinguish between the two cases in eq.~(\ref{dmns-pair}), 
which give the same $\sin\dmns$, we should 
measure $\cos\dmns$ as well.  
In the $\nu_\mu \to \nu_e$ transition probability expression of 
eq.~(\ref{pnumu2nue}), the relevant term is 
\bea
\label{cos-dmns}
2Re[U_{\mu 2}U^*_{e2}U_{e3}U^*_{\mu3}]\sin\Delta_{13}\cdot\Delta_{12} \,,
\eea
whose magnitude is small at around $\Delta_{13} \sim \pi$ where 
the leading $\sin^2 (\Delta_{13}/2)$ term takes its maximum value 
of unity.  
In order to distinguish this sub-leading term from the leading one, 
we need to make experiments at different neutrino energies.  
By choosing the neutrino narrow band beam at two different energies, 
we could show in ref.~\cite{t2hk} that the degeneracy between  
$\dmns=0^\circ$ and $180^\circ$ can be resolved.    

After our work \cite{t2hk} was completed in 2002, 
the T2K collaboration decided to adopt the off-axis beam (OAB) 
rather than the narrow band beam (NBB) which was used in our analysis. 
The OAB flux has a sharp peak at low energies and has relatively 
long high-energy tail.  The flux intensity of the OAB around the sharp 
peak is typically stronger than that of the NBB, while the hard high 
energy tail may be considered as a disadvantage.  
We show the OAB flux profile from the J-PARC 50~GeV proton 
accelerator in Fig.~\ref{fig:oab}.  Fluxes at off-axis angles 
between $0^\circ$ (on-axis) and $3^\circ$ are shown for later 
use.  For the T2K and T-to-HK experiments, the neutrino beam 
line at J-PARC has been set up such that the OAB beam between 
$2^\circ$ and $3^\circ$ can be sent to SK and HK.  
The beam orientation has been so chosen that once HK is constructed 
at its planned location, exactly the same OAB will hit both detectors.  
We may read off from the figure that the peak flux energy is about 
0.8, 0.7, and 0.6~GeV for $2^\circ$, $2.5^\circ$, and $3^\circ$, 
respectively.  The larger the OAB angle, the lower the peak 
energy is, and sharper the cut-off for the high-energy tail.    

\begin{figure}[p]
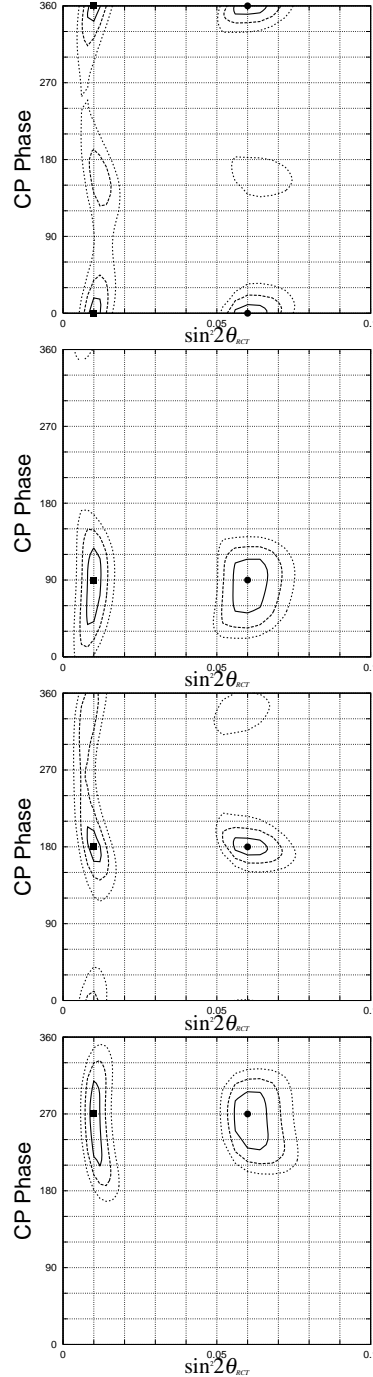
%[htbp]
\begin{center}
\includegraphics[angle=-90,width=4.8cm]{t2h_oab_000.eps}
\includegraphics[angle=-90,width=4.8cm]{t2h_oab_090.eps}
\includegraphics[angle=-90,width=4.8cm]{t2h_oab_180.eps}
\includegraphics[angle=-90,width=4.8cm]{t2h_oab_270.eps}
\caption{%
Allowed regions in the plane of $\sin^22\theta_{_{\rm RCT}}$
and $\delta_{_{\rm MNS}}$ after 6 years of T-to-HK.  
}
\label{fig:t2hk}
\end{center}
\end{figure}

By repeating the analysis of ref.~\cite{t2hk} for the experimental 
setup of 
\bseq
\bea
&&\mbox{\rm 0.8 Mton}\cdot\mbox{\rm year for}\; 
\nu_\mu \mbox{\rm OAB}(2^\circ) \,,  
\\
&&\mbox{\rm 2.0 Mton}\cdot\mbox{\rm year for}\; 
\nu_\mu \mbox{\rm OAB}(3^\circ) \,,  
\\
&&\mbox{\rm 3.2 Mton}\cdot\mbox{\rm year for}\; 
\ov\nu_\mu \mbox{\rm OAB}(2^\circ) \,,  
\eea
\eseq
we find the results of Fig.~\ref{fig:t2hk}.  
The flux ratios are so chosen that roughly the same number of charged 
current events are expected, about $10^4$ $\mu$-like events 
and about a few hundreds to a thousand $e$-like events, in the range 
of the parameter space we studied.  
If a factor of 5 more intensive beam is available at J-PARC by 
the time we will have HK, the same quality results may be obtained 
in less than one and a half year time.  

The contours of Fig.~\ref{fig:t2hk} show the 1-, 2-, and 3-$\sigma$ 
allowed region in the plane of $\sin^2 2\thrct$ and $\dmns$.  
They are obtained as follows.  
We first calculate the expected numbers of $\mu$-like and $e$-like 
events at HK for the following input parameter values:  
\bseq
\label{t2hk-inputs}
\bea
\delta m^{2}_{_{\rm ATM}}&=& 3 \times10^{-3} ~{\rm eV}^2 \,,
\\
\delta m^{2}_{_{\rm SOL}}&=& 7\times10^{-5} ~{\rm eV}^2 \,,
\\
\sin^2 2\theta_{_{\rm ATM}}&=& 1 \,,
\label{thatm0.5}
\\
\sin^2 2\theta_{_{\rm SOL}}&=& 0.85 \,,
\\
\sin^2 2\theta_{_{\rm RCT}}&=& 0.01 \;\mbox{\rm or}\; 0.06 \,,
\\
\delta_{_{\rm MNS}} &=& 0^\circ \,,90^\circ \,,180^\circ 
\;\mbox{\rm or}\; 270^\circ \,,\quad
\eea
\eseq
by assuming the constant matter density of $\rho=3$g/(cm)$^3$ 
along the baseline.  
The resulting number of events are then analyzed by varying 
all the 6 parameters under the known constraints for the four 
parameters $\dmsqatm$, $\sin^2 2\thatm$, $\dmsqsol$, and 
$\sin^2 2\thsol=4\tan^2\thsol/(1+\tan^2\thsol)^2$, 
and by allowing for 3\% normalization uncertainties in the 
earth matter density and the overall beam flux.  
In addition, some of the known systematic errors from the 
SK and K2K experiences were taken into account.  

We find from Fig.~\ref{fig:t2hk} that the T-to-HK experiment can 
establish CP violation at the 3-$\sigma$ level if 
$\dmns \sim 90^\circ$ or $270^\circ$.  
Also, the use of both $2^\circ$ and $3^\circ$ OAB can contribute 
to resolving the degeneracy between 
$\dmns \sim 0^\circ$ or $180^\circ$ if 
$\sin^2 2\thrct$ is not too small.  
When comparing the contours of Fig.~\ref{fig:t2hk} with the 
corresponding results of ref.~\cite{t2hk} which were obtained 
for the narrow-band beams (NBB), however, we find that the 
areas of the 3-$\sigma$ allowed regions are significantly wider 
for the OAB than the NBB and that the capability of distinguishing 
between the $0^\circ$ and $180^\circ$ degeneracy is lower for 
the OAB.  
These results are obtained even though the typical flux of OAB 
is higher than that of NBB, mainly because of much higher background 
of the OAB, especially in the high energy tail.  
The background is found most serious for the $\ov\nu_\mu \to 
\ov\nu_e$ measurements where the $\nu_e$ component of the 
$\ov\nu_\mu$ enriched beam contributes to the $e$-like signal 
since HK cannot distinguish the charge.  
We find that the background level is significant in the high energy 
tail of the OAB, and hence rejection of events with high neutrino 
energies can improve the measurement.  
Our studies on the T-to-HK experiments with OAB will be reported 
in \cite{t2hk-oab} soon.  

\section{The parameter degeneracy problem} 

Those studies, however, assumed that the neutrino mass hierarchy is
known to be normal $(\delta m^2_{13}=\matm{})$ and also we assumed
$\satms{}=0.5$ as an input.

In particular, our results for $\dmns = 90^\circ$ and $270^\circ$ 
in Fig.~\ref{fig:t2hk} suggest that the T-to-HK experiment can 
establish CP violation at the 3-$\sigma$ level or higher, if 
$\sin^2 2\thrct$ is not too small.  
We find, however, that this observation depends critically on 
the assumption that we know the neutrino mass hierarchy to be 
normal, i.e., the hierarchy pattern I in Fig.~\ref{fig:hi-1234}. 

\begin{figure}[t]
\includegraphics[width=6cm]{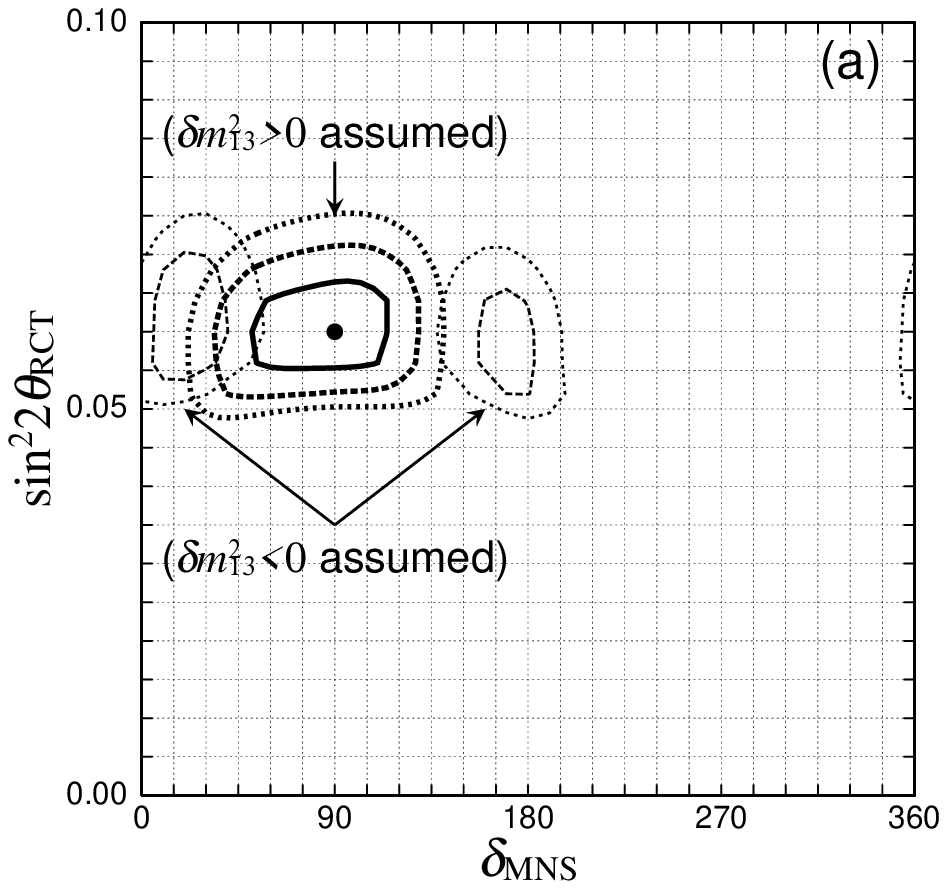}
\includegraphics[width=6cm]{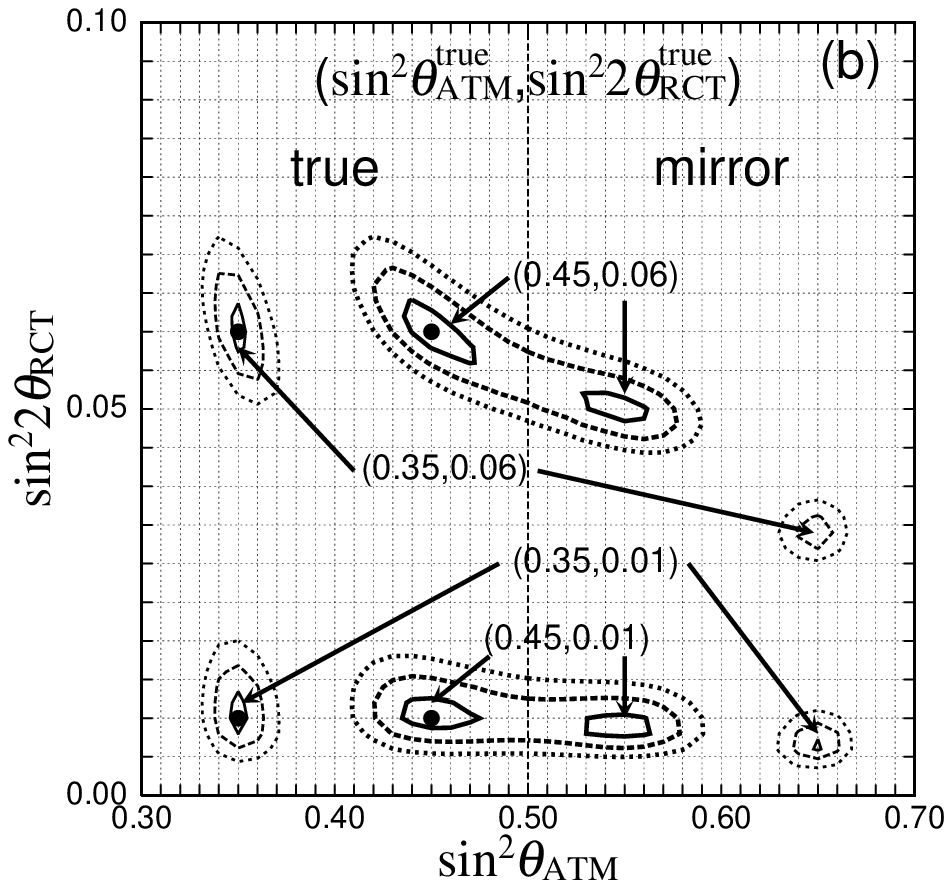}
\caption{%
Doubling of the regions allowed by future LBL experiment T-to-HK, 
between J-PARC at Tokai and HyperKamiokande, in the plane of 
$\delta_{_{\rm MNS}}$ and $\sin^22\theta_{_{\rm RCT}}$ (a),
and $\sin^2\theta_{_{\rm ATM}}$ and $\sin^22\theta_{_{\rm RCT}}$ (b).
}
\label{fig:t2hk-degeneracy}
\end{figure}

As a demonstration, we show in Fig.~\ref{fig:t2hk-degeneracy}(a)  
the allowed region of the T-to-HK simulation by analyzing 
exactly the same input `data', calculated for the normal 
hierarchy ($\delta m^2_{13} > 0$), by assuming that the neutrino 
mass hierarchy is inverted ($\delta m^2_{13} < 0$).  
We find the region encircled by thin lines from this analysis.  
The minimum of the $\chi^2$ value in those regions are only 
slightly larger than unity, and hence the solutions cannot 
be excluded by the measurements.  
Most strikingly, the allowed regions obtained by assuming 
the inverted hierarchy give $\dmns \sim 0^\circ$ or $180^\circ$, 
i.e., the very conclusion of the evidence of CP violation in 
the lepton sector depends on our capability of distinguishing 
between the two remaining hierarchy cases, I (normal) or 
III (inverted).

There is one more hidden degeneracy problem among the 6 parameters 
of the 3 neutrino model.  
It is the magnitude of $\sin^2 \thatm=|U_{\mu 3}|^2$ which has 
a two-fold ambiguity 
\bea
\label{thatm-pm}
\sin^2 \thatm = |U_{\mu 3}|^2 = 
\frac{1\pm\sqrt{1-\sin^2 2\thatm}}{2} \quad
\eea
in the last line of eq.~(\ref{3angles}).  
The two-fold ambiguity was hidden so far because we chose the 
extreme value of $\sin^2 2\thatm=1$ as our input.  
It should be remarked that the presently allowed range of 
$\sin^2 2\thatm>0.90$ in eq.~(\ref{data-atm}) corresponds to 
\bea
\label{thatm-dat}
0.34 < \sin^2\thatm < 0.65 \,.
\eea
In Fig.~\ref{fig:t2hk-degeneracy}(b), we showed the allowed 
region in the same T-to-HK simulation, when the input data 
are calculated for 
\bea
\label{t2hk-input2}
\sin^2 \thatm = 0.35 \;\mbox{\rm or}\; 0.45 
\eea
instead of $\sin^2 \thatm=0.5$ in eq.~(\ref{t2hk-inputs}).  
In addition to the correct solutions which are marked by the 
solid blob at the input points, we find a mirror solution 
at the $\sin^2 \thatm$ value corresponding to the opposite 
sign in eq.~(\ref{thatm-pm}) for the same $\sin^2 2\thatm$ value.
We notice also that the mirror solutions appear at different  
values of $\sin^2 2\thrct$, and hence the measurement of 
$\sin^2 2\thrct$ at T2K and in the future T-to-HK experiments 
also suffer from this degeneracy problem.  
The reason for the correlation between $\sin^2 \thatm$ and 
$\sin^2 2\thrct$ is a manifestation of the fact that the 
magnitude of the leading term that dictates the 
$\nu_\mu \to \nu_e$ oscillation probability (\ref{pnumu2nue}) 
is their product in eq.~(\ref{numu2nuefactor}).  
%Since the mirror solution for $\sin^2\thatm$ is greater than 
%its true value, the corresponding preferred value of 
%$\sin^22\thrct$ is smaller.  
This correlation suggests that if we can measure 
$\sin^2 2\thrct$ that governs the leading term of the 
$\nu_e \to \nu_e$ oscillation in eq.~(\ref{pnue2nue}) 
accurately independent of the magnitude of $\sin^2\thatm$, 
we can resolve the ambiguity.   

These studies demonstrate clearly that we need more works 
to do in order to determine the neutrino mass hierarchy pattern 
between the normal (I) and inverted (III) ones, and to determine 
the third mixing angle $\sin^2 2\thrct$ and $\sin^2 \thatm$ 
independently.  
New generations of the reactor neutrino experiments, 
such as Double-CHOOZ \cite{chooz2} and KASKA \cite{kaska} 
have been proposed to tackle the latter problem.  
In the rest of this talk, I would like to introduce our studies 
about possible future long baseline neutrino oscillation experiments 
in Asia, which make use of neutrino beams from the J-PARC facility.  

\section{Fate of the T2K and T-to-HK beam}  

We first study the fate of the off axis beam of the T2K and the 
possible T-to-HK project.    
The OAB at J-PARC is shoot underground, such that exactly the same 
OAB will be available at the present SK site and the planned HK site.  
The center of the beam at around Kamioka site passes through more than 
10~km deep in the earth crust.   
The beam center will re-appear on the earth surface in the middle 
of the Japan sea, and the OAB of the off-axis angle between 
$0.5^\circ$ and $3^\circ$ will be available free in Korea.  
The surface view of the OAB($2.5^\circ$) is shown in 
%%%%%%%%%%%%%%%%%%%%%%%%%%%%%%%%%%%%%%%%%%%%%%%%%%%%%%%%%%%%%%%
Fig.~8 %Fig.~\ref{t2kr:h-view} 
%%%%%%%%%%%%%%%%%%%%%%%%%%%%%%%%%%%%%%%%%%%%%%%%%%%%%%%%%%%%%%%
and the vertical cross section view 
of the beam is shown in 
%%%%%%%%%%%%%%%%%%%%%%%%%%%%%%%%%%%%%%%%%%%%%%%%%%%%%%%%%%%%%%%
Fig.~9. %Fig.~\ref{t2kr:v-view}.  
%%%%%%%%%%%%%%%%%%%%%%%%%%%%%%%%%%%%%%%%%%%%%%%%%%%%%%%%%%%%%%%

In the surface view of 
%%%%%%%%%%%%%%%%%%%%%%%%%%%%%%%%%%%%%%%%%%%%%%%%%%%%%%%%%%%%%%%
Fig.~8, %Fig.~\ref{t2kr:h-view}, 
%%%%%%%%%%%%%%%%%%%%%%%%%%%%%%%%%%%%%%%%%%%%%%%%%%%%%%%%%%%%%%%
the red curves show the contours of the same baseline length.  
The solid and dashed black contours give the off-axis angle.  
The center of the beam is at $0^\circ$, which appears in the 
Japan sea, a few hundred km east from the Korean coast.   
Inside of Korea mainland, the off-axis beam of $1^\circ$ to 
$3^\circ$ is available at $L=$1,000~km to 1,200~km.  
The off axis beam at $0.5^\circ$ is available at the Korean 
east coast, for the OAB($3^\circ$) at Kamioka.  

Therefore, depending on the location of a possible neutrino 
detector in Korea, the flux of off-axis angle between 
$0.5^\circ$ and $3^\circ$ shown in Fig.~\ref{fig:oab} 
will be available during the period of the T2K experiment, 
and possibly during the T-to-HK experiment.  
Because the flux at $L\sim$ 1,000~km to 1,200~km is about 
a factor of 10 to 16 smaller than that at Kamioka, respectively, 
we may need a detector of 100~kton volume in order to obtain  
significant new contributions.  

Because the OAB at large angles has a sharp peak at low energies, 
see Fig.~\ref{fig:oab}, one may study the oscillation 
probabilities at large $L/E$.  
There are a few advantages of neutrino oscillation experiments 
at large $L/E$ while keeping the neutrino energy at sub-GeV.  
\begin{itemize}
\item{
$\Delta_{12}$ grows with $L/E$, and hence the sensitivity 
to $\dmns$ in the $\nu_\mu \to \nu_e$ experiment will grow; 
see eq.~(\ref{pnumu2nue}). }
\item{
The sign of the term linear in both $\Delta_{12}$ and 
$\Delta_{13}$ determines the neutrino mass hierarchy, 
because the term $\Delta_{12}\cdot \Delta_{13}$ is positive 
for the normal hierarchy (I), while it is negative for the 
inverted hierarchy (III).} 
\end{itemize}
Relative smallness of the earth matter effects at lower energies 
will be helpful in the study, since the above features 
of the $\nu_\mu \to \nu_e$ oscillation probability can be read 
off from the expansion in terms of $\Delta_{12}$ in 
eq.~(\ref{pnumu2nue}) which is a good approximation only 
up to a few GeV; see Fig.~\ref{fig:matter}.  

\begin{figure}[p]
\begin{center}
\includegraphics[width=5.8cm]{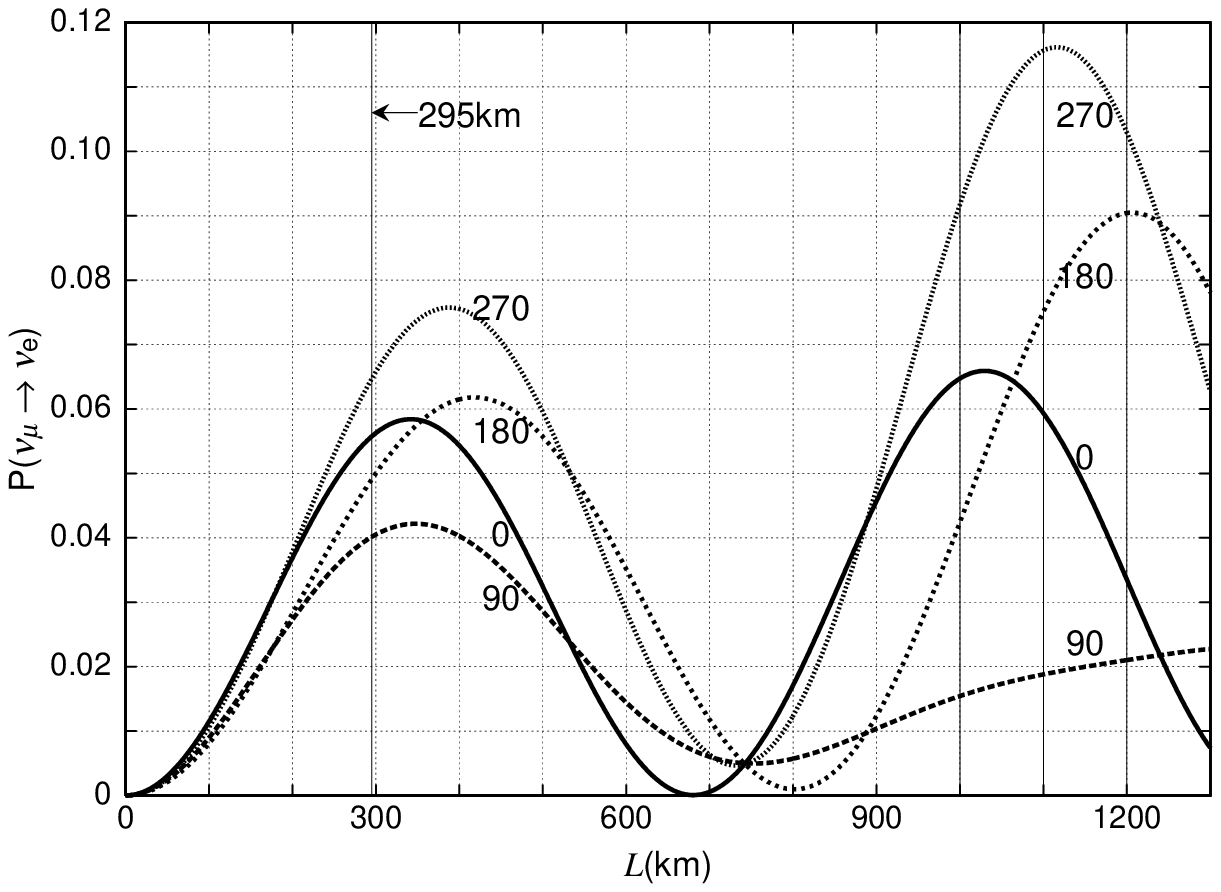}
\includegraphics[width=5.8cm]{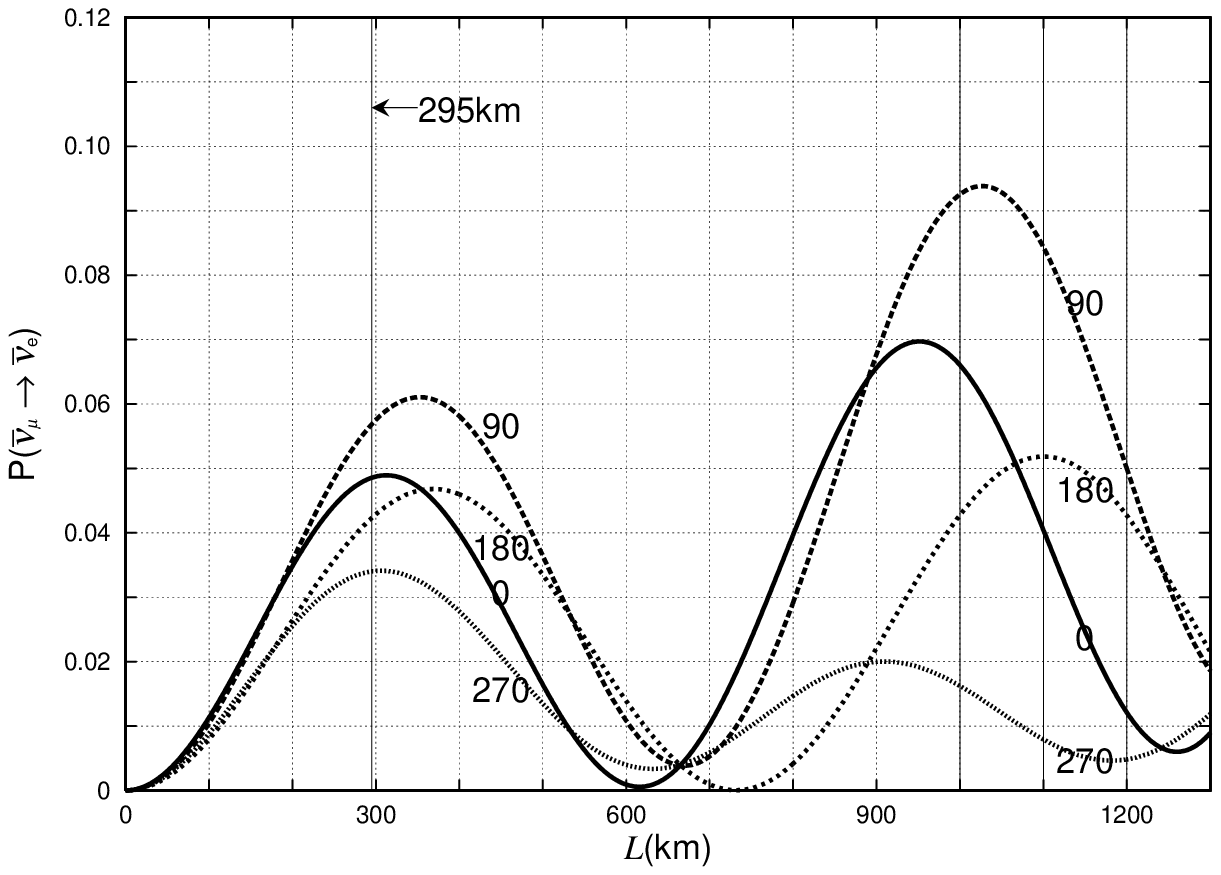}
\includegraphics[width=5.8cm]{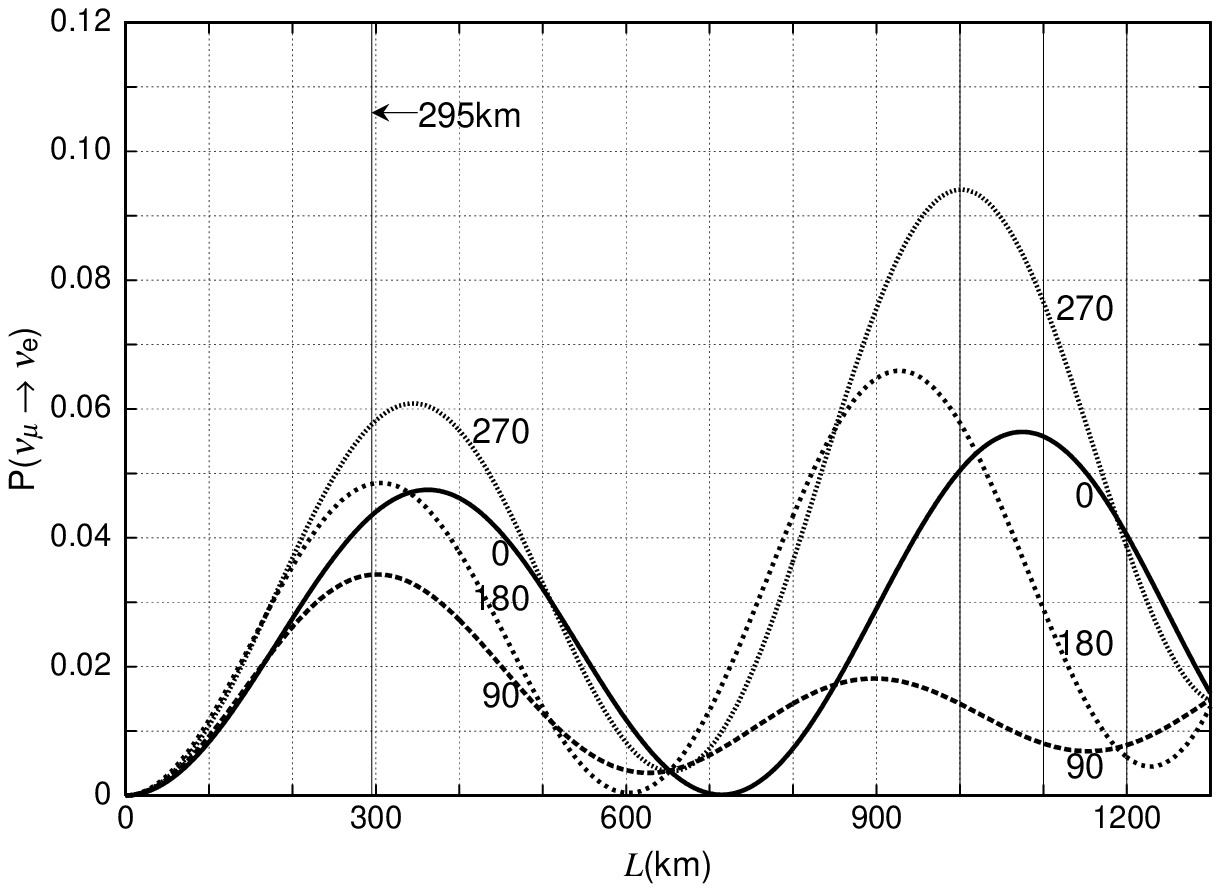}
\includegraphics[width=5.8cm]{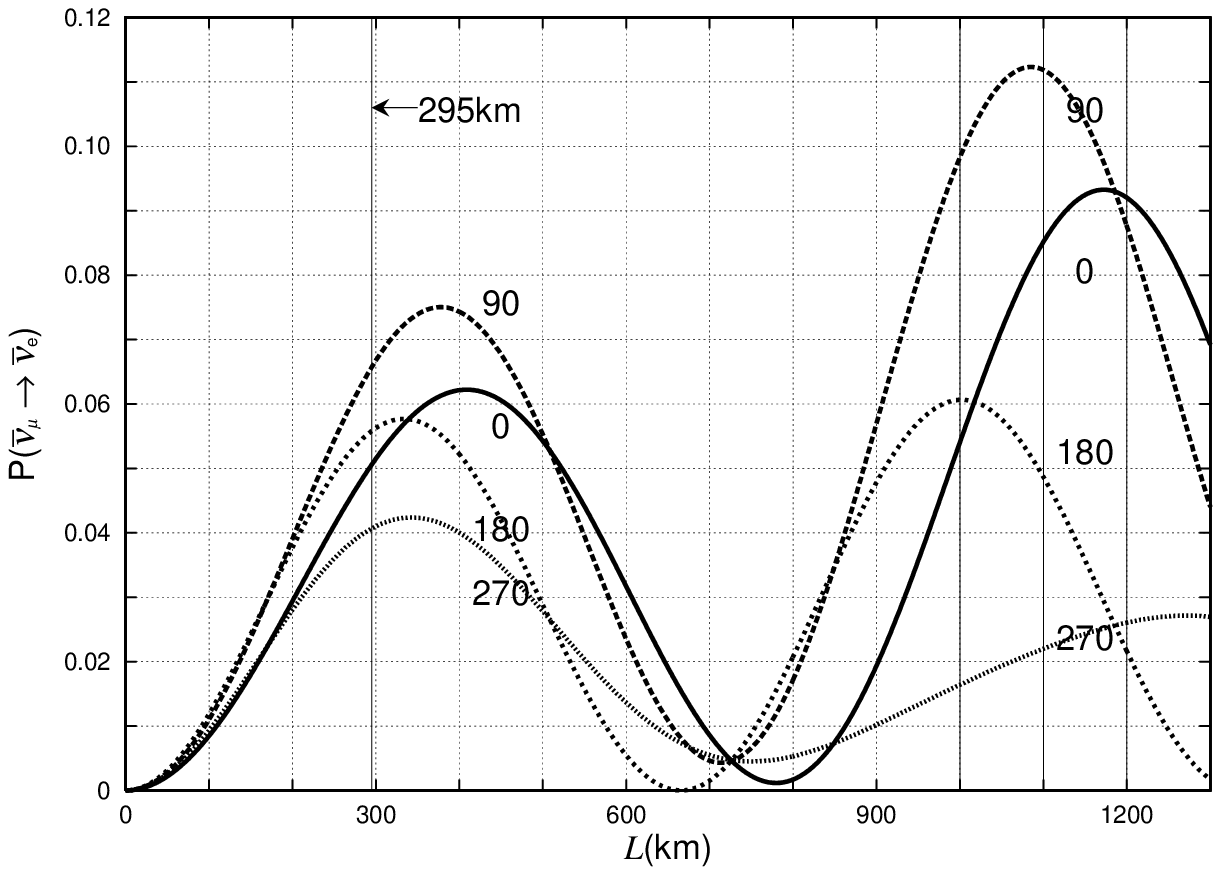}
\end{center}
\caption{%
The oscillation probabilities $P(\nu_\mu \to \nu_e)$ and 
$P(\ov\nu_\mu \to \ov\nu_e)$ for $E_\nu=0.7$~GeV plotted 
against the baseline length $L$.  
The upper two figures are for the normal hierarchy (I) 
and the lower two are for the inverted hierarchy (III).  
}
\label{fig:t2kr_1vs3}
\end{figure}

In Fig.~\ref{fig:t2kr_1vs3}, we show the oscillation probabilities 
$P(\nu_\mu \to \nu_e)$ and $P(\ov\nu_\mu \to \ov\nu_e)$ for 
$E_\nu=0.7$~GeV plotted against the baseline length $L$.  
The upper two figures are for the normal hierarchy (I) 
and the lower two are for the inverted hierarchy (III).  
We can see from the figure that in some cases, small differences 
at $L=300$~km for the T2K and T-to-HK experiments can be enhanced 
significantly in Korea between $L=$1,000~km and 1,200~km.  
The difference between the normal and inverted hierarchy can 
be very significant and there is a hope that a combination of 
the T-to-HK measurement and the T-to-Korea experiment can 
resolve the ambiguity in the $\dmns$ and the neutrino mass 
hierarchy cases at the same time.  
The study is underway \cite{t2kr}.  

%\begin{figure*}[p]
\begin{figure*}[t]
\begin{center}
\includegraphics[width=16cm]{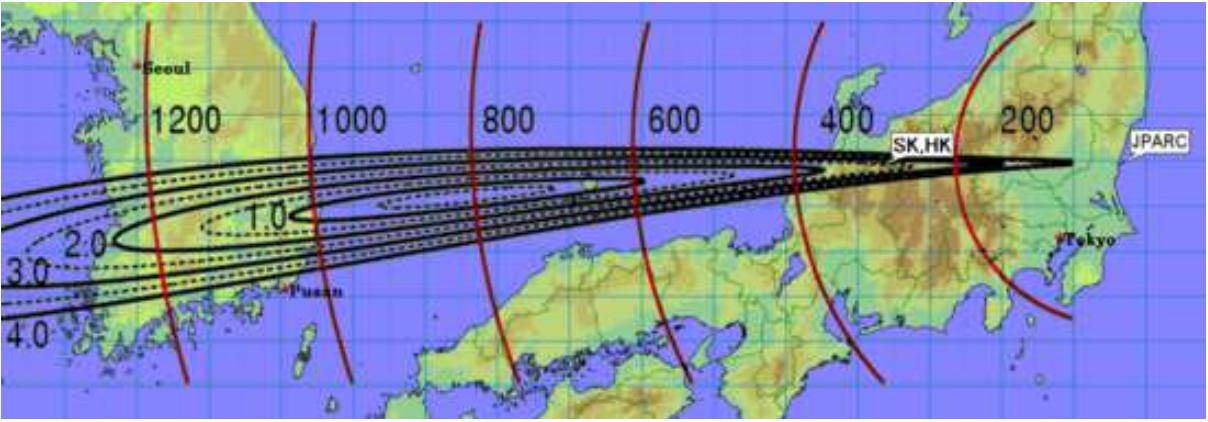}
\caption{%
The fate of the OAB 2.5 degree beam from J-PARC.  Surface view. 
}
\end{center}
\label{t2kr:h-view}
\end{figure*}

\begin{figure}[h]%[htbp]
\begin{center}
\includegraphics[width=8cm]{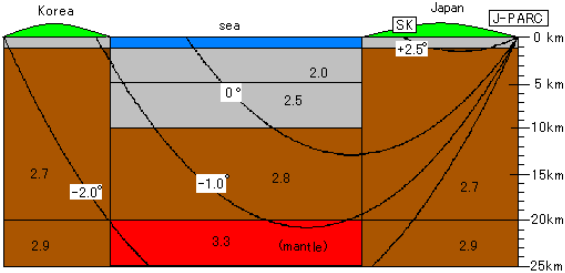}
\caption{%
The fate of the OAB 2.5 degree beam from J-PARC.  Vertical view.  
}
\end{center}
\label{t2kr:v-view}
\end{figure}

At the moment, we do not know if there is a strong interest in 
constructing a huge neutrino detector in Korea.  
Because we feel that this investigation is rather promising, we study 
the fate of the OAB from J-PARC in the T2K experiment carefully.  
Although only the fate of the OAB($2.5^\circ$) is shown in 
%%%%%%%%%%%%%%%%%%%%%%%%%%%%%%%%%%%%%%%%%%%%%%%%%%%%%%%%%%%%%%%
Fig.~8 and Fig.~9, %Fig.~\ref{t2kr:h-view} and Fig.~\ref{t2kr:v-view}, 
%%%%%%%%%%%%%%%%%%%%%%%%%%%%%%%%%%%%%%%%%%%%%%%%%%%%%%%%%%%%%%%
we made similar figures for OAB($2^\circ$) and OAB($3^\circ$).  
In case of OAB($2^\circ$), the smallest off-axis angle at the 
east coast of Korea is about $1.5^\circ$, while for OAB($3^\circ$), 
the smallest angle is about $0.5^\circ$.  

In the cross section view of 
%%%%%%%%%%%%%%%%%%%%%%%%%%%%%%%%%%%%%%%%%%%%%%%%%%%%%%%%%%%%%%%
Fig.~9, %Fig.~\ref{t2kr:v-view}, 
%%%%%%%%%%%%%%%%%%%%%%%%%%%%%%%%%%%%%%%%%%%%%%%%%%%%%%%%%%%%%%%
we show the baseline, along the direction of J-PARC at Tokai village 
to the center of the SK and the proposed HK site in Kamioka.  
The OAB($2.5^\circ$) beam hits SK and HK at the same off-axis 
angle of $2.5^\circ$, while the beam center ($0^\circ$) reaches 
the earth surface in the sea.  It is rather important to recognize 
that the Japan sea is rather shallow, and the beams which reach 
Korea do not travel under water.  The numerics in the earth 
crust show the effective mass density $\rho$ in units of 
g/(cm)$^3$.  We can view that under the sea the earth crust 
is thin, and the mantle appears already at 20~km beneath the 
earth surface.  The density variations, however, are not 
expected to be significant.  
We hope that our studies \cite{t2kr} will be found useful 
by our colleagues in Korea.  

\section{High energy super beam to China}  

%\begin{figure}[p]
%\includegraphics[width=16cm]{t2b_view.eps}
\begin{figure}[t]
\includegraphics[width=7.5cm]{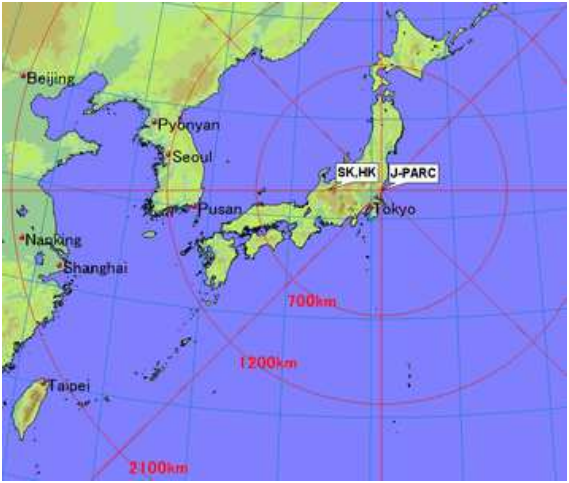}
\caption{%
East Asia viewed from J-PARC in Tokai.  
The baseline length $L=$~295~km (Kamioka), 1,200~km (Seoul) and 
2,100~km (Beijing).  
}
\label{t2b-view}
\end{figure}
%\end{figure*}

\begin{figure}[t]
\includegraphics[width=8cm]{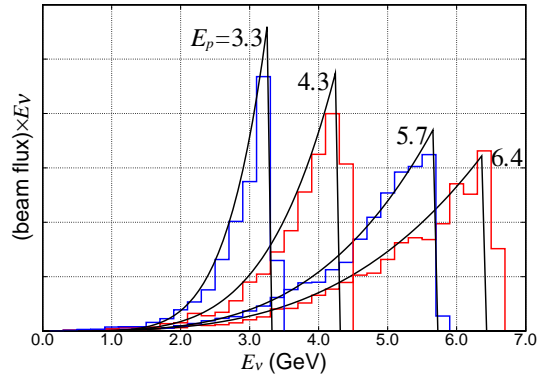}
\caption{%
High energy narrow band beams (HENBB) at J-PARC 
for the possible T2B (Tokai-to-Beijing) project.  
}
\label{t2b-flux}
\end{figure}

\begin{figure}[ht]
\includegraphics[width=8cm]{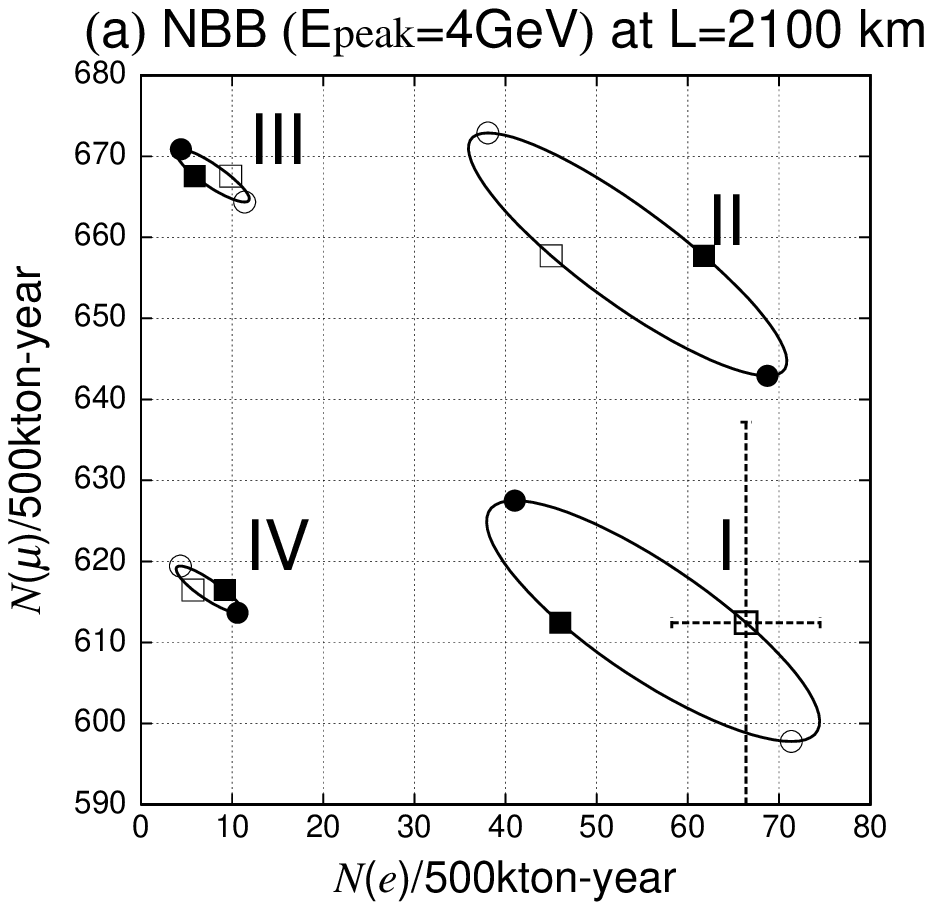}
\caption{%
The expected number of mu-like and e-like events for the future 
T2B experiment with the beam HENBB(4GeV).
}
\label{t2b-1234}
\end{figure}

Although we do not yet know if there are strong enough interests in 
constructing a huge neutrino detector in Korea, 
strong interests have been expressed by our Chinese colleagues 
about the possibility of sending super neutrino beams from J-PARC 
at Tokai to somewhere in mainland China.  
A possible 100~kton level water \cerenkov detector BAND 
(Beijing Astrophysics and Neutrino Detector) \cite{BAND} 
has been proposed, and if it will be placed in Beijing, the baseline 
length from Tokai will be about $L$=2,100~km.  
The unique capability of the BAND detector is that it is a segmented 
detector, where each segment contains 10~ton of water surrounded 
by the photo-multiplier tube to measure the \cerenkov light, and 
at the same time has a good capability of calorimetric 
energy measurement \cite{BAND}.

Encouraged by the interests expressed by our Chinese colleagues, 
we made a serious study of physics capability of 
a very long baseline neutrino oscillation experiments between 
J-PARC at Tokai and BAND in Beijing \cite{t2b}.  
The global view of the eastern Asia seen from J-PARC at Tokai 
is shown in Fig.~\ref{t2b-view}.  
It is clearly seen from the figure that the LBL experiments 
with L greater than 2,000~km are feasible 
if a huge neutrino detector can be built in China.  
We can also tell from the figure that the direction from Tokai 
to Kamioka is almost exactly from east to west, and the T2K 
beam goes through southern part of Korea.  
Since Beijing is placed north east from the T2K beam-line, 
a new neutrino beam line toward Beijing should be constructed 
at J-PARC.  

In addition, in order to send sufficient neutrino flux to the 
distance greater than 2,000~km, and in order to study physics 
at the same $L/E$ region where the first oscillation mode takes 
its maximum value, 
\bea
\label{l-ov-e}
|\Delta_{13}| \sim \pi \;\;\mbox{\rm when}\;\; 
L(\mbox{\rm km})/E(\mbox{\rm GeV}) \sim 500 \,, 
\eea
for $\dmsqatm = 2.5\times 10^{-3}$~eV$^2$, 
we need a neutrino beam whose flux peaks at around 3 to 6 GeV.  
Such neutrino beams can indeed be designed for the J-PARC 
50~GeV proton accelerator, and has been named HENBB (High 
Energy Narrow Band Beam) in ref.~\cite{t2b}.  
We show in Fig.~\ref{t2b-flux} the expected flux times the neutrino 
energy $E_\nu$ as functions of $E_\nu$.  
We multiply the flux by $E_\nu$ in order to show the peak 
in the expected number of events in the absence of the 
neutrino oscillation.  
The peak position of the flux $\times E_\nu$ is called $E_p$ 
and the simulation results are shown for 
$E_p =$ 3.3, 4.3, 5.7 and 6.4~GeV.  
The optics adopted for generating the HENBB makes the peak rather 
sharp, especially in the high energy side.  
This sharp cut-off of the high energy neutrinos makes the 
background level low when the HENBB is used for the LBL experiments.  
In fact we find that the largest background from the high energy 
neutrinos beyond $E_p$ comes mainly from the contribution from 
the $K \to \mu \nu$ decays, which may be measured precisely at  
a near detector.  
We made a parameterization of the HENBB flux for a given $E_p$ 
in order to save time for the beam simulation, and the 
parameterization is given in ref.~\cite{t2b}.  

We show in Fig.~\ref{t2b-1234}, the expected total numbers of 
the $e$-like events and the $\mu$-like events.  
Since we do not require the charge identification capability 
for the BAND detector\footnote{
Possibility of putting the segmented water \cerenkov detector 
of BAND under the magnetic field has been studied by the IHEP 
group \cite{BAND}.}
the $\mu$-like events are the sum of $\nu_\mu$ and $\ov\nu_\mu$ 
charged current (CC) events and the $e$-like events are the sum of 
$\nu_e$ and $\ov\nu_e$ CC events and those expected from the 
NC events where produced $\pi^0$'s are mistaken as $e^\pm$.  
At higher energies, $\tau$ production from the dominant 
$\nu_\mu \to \nu_\tau$ oscillation mode becomes significant, 
and their leptonic decay modes have also been counted as 
background.  
The contours shown in Fig.~\ref{t2b-1234} are the prediction 
of the three neutrino model for 
$\dmsqatm = 3.5\times 10^{-3}$~eV$^2$, which was the preferred 
value a few years ago.  
The results are for 500~kton$\cdot$year, or for 5 years with 
the 0.75~MW J-PARC and a 100~kton BAND.  
They are shown for all the hierarchy cases I, II, III, IV 
of Fig.~\ref{fig:hi-1234}, where the results for the cases II and IV 
can be interpreted as the expectation for anti-neutrino oscillation 
experiments, according to the theorem eq.~(\ref{hi-theorem}), 
after the ratio 
$\sigma_{\cc}^{}(\ov\nu_\ell)
/\sigma_{\cc}^{}(\nu_\ell) \sim 1/3$ 
is multiplied,.  

It is most striking to learn from the figure that the 
$\nu_\mu \to \nu_e$ oscillation probability is almost a factor of 
7 smaller in the inverted hierarchy (III) than the normal hierarchy (I).  
On the contrary, the $\ov\nu_\mu \to \ov\nu_e$ oscillation 
is suppressed for the normal hierarchy (IV).  
This is clearly due to the enhanced matter effects at high energies, 
because such significant suppression has not been identified in our 
previous studies for the T-to-HK project, where the observation of 
the $\ov\nu_\mu \to \ov\nu_e$ events are essential to determine 
the CP-violating parameter, $\jmns$ or $\sin\dmns$.  
The two experiments measure approximately the same phase of the 
oscillation probability, because $L/E = 295~$km$/$0.7~GeV is 
not far from 2,100~km$/$4~GeV.  

This can be understood qualitatively as follows.  
The oscillation probability $P_{\nu_\mu \to \nu_e}$ inside of the matter 
can be expressed as eq.~(\ref{pnumu2nue}), where the mass squared 
differences and the mixing matrix are replaced by the corresponding 
ones $\delta\wt{m}^2_{ij}$ and $\wt{U}_{\alpha i}$ according to 
eq.~(\ref{replacement}).  
From eq.~(\ref{pnumu2nue}), we can tell that the difference in 
$P_{\nu_\mu \to \nu_e}$ between the mass hierarchy I and III 
should come from the difference in the second term proportional to 
$\Delta_{12}$, which changes the sign between I and III.  
We find that the mixing matrix 
terms do not change significantly in the relevant energy region, 
and the second term contributes constructively for the normal 
hierarchy (I), whereas destructively for the inverted case (III), 
as long as  $|\wt\Delta_{13}|<\pi$.  
From Fig.~\ref{fig:matter}, we find that the magnitude of the 
term $\delta\wt{m}^2_{12}$ grows rapidly in a few to several GeV region, 
approaching that of the leading terms of the order $\dmsqatm$.  
Although the similar trend is observed for the oscillation at 
$E_\nu \sim 0.7$~GeV for the T2K and T-to-HK projects, 
the effects are small enough to be compensated by shifts 
in the other model parameters, notably $\sin^2 2\thrct$ and $\dmns$.  

\begin{figure}[t]
\includegraphics[width=7cm]{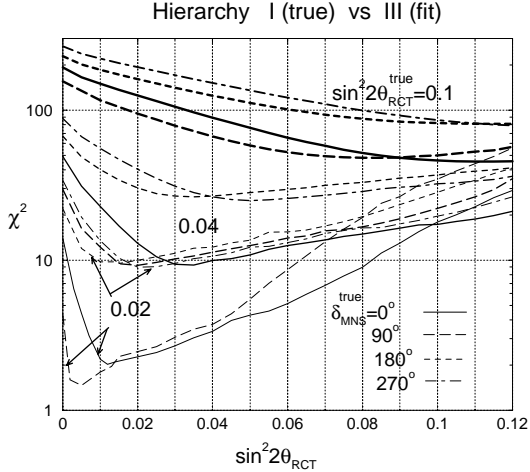}
\caption{%
The expected minimum chi-squared of the combined T2K and T2B 
experiments when the normal hierarchy is realized in the nature 
while the inverted hierarchy is wrongly assumed in the analysis.   
}
\label{t2b-1vs3}
\end{figure}

This drastic suppression of $\nu_\mu \to \nu_e$ oscillation 
probability for the inverted hierarchy at high energies gives 
us a powerful tool to determine the neutrino mass hierarchy at 
very long baseline neutrino oscillation experiments 
which are capable of detecting electrons.  
The difference in the predictions of the normal and the inverted 
hierarchy cases at $\sin^2 2\thrct \sim 0.1$ is so huge that 
no adjustment of the other model parameters can compensate for it.  
The Tokai-to-Beijing project with HENBB seems to be an ideal example 
of such experiments.  
In Fig.~\ref{t2b-1vs3}, we show the results of simple statistical 
analysis \cite{t2b}, where we calculated the number of expected 
events at BAND after 5 years of HENBB at $E_p=4$~GeV and another 
5 years at 6~GeV for the normal hierarchy and then by 
analyzing the `data' by assuming the inverted hierarchy.  
The results of our analysis depend seriously on the assumed value 
of $\sin^2 2\thrct$, and hence the \chisqmin values 
of our statistical analysis is shown as functions of the  
$\sin^2 2\thrct$ value in the fit.  
We examined the cases with the input (`true') parameters at 
$\sin^2 2\theta_{\rct}^{\true}=$ 0.1, 0.04 and 0.02, for the 4 phase angles, 
$\delta_{\mns}^{\true} = 0^\circ, 90^\circ, 180^\circ$ and $270^\circ$.  
It can be seen from the figure that if 
$\sin^2 2\theta_{\rct}^{\true} > 0.04$ 
and the hierarchy is normal (I), then the inverted hierarchy case 
(III) can be excluded at 3-$\sigma$ level for any values of $\dmns$.  
On the other hand, the generated `data' at 
$\sin^2 2\theta_{\rct}^{\true} = 0.02$ 
for $\delta_{\mns}^{\true} = 0^\circ$ and $90^\circ$ can be 
fitted well within the allowed range of the other model parameters.  

We also studied the possibility of constraining 
the $\dmns$ parameter in the possible Tokai-to-Beijing project by 
using the HENBB, but the results we find is not as promising 
as those from the analysis \cite{t2hk} of the possible Tokai-to-HK 
project.  The reason is rather simple.  
If the neutrino mass hierarchy is normal, 
the oscillation probability for the $\ov\nu_\mu \to \ov\nu_e$ 
is strongly suppressed, and hence we cannot expect much number of 
events by sending anti-neutrino beams.  
If the hierarchy is inverted, on the other hand, the 
$\nu_\mu \to \nu_e$ oscillation probability is strongly suppressed, 
and we will need significantly higher power neutrino beams.  
If the inverted hierarchy is realized in the nature, the 
Tokai-to-Beijing project with a 4~MW proton accelerator at J-PARC 
may be necessary.  
In the Tokai-to-HK project, the use of the low energy neutrino beam 
helps reducing of the matter effects and a one Mton detector 
more than compensates for the small cross section at low energies.  

In ref.~\cite{t2b}, we show the results of our analysis at 
$L=1,200$km with the HENBB at $E_p=3$~GeV and 5~GeV, and find 
nearly as promising results as the above Tokai-to-Beijing analysis.  
If there appears a strong interest in constructing a huge neutrino 
detector in Korea, this might as well be an appealing option.  
There, the remnant of the OAB will be studied during the period of 
the T2K and the planned T-to-HK project, and at a later stage, 
a new beam-line for the HENBB may be constructed at the J-PARC 
site to determine the neutrino mass hierarchy.  

\section{Neutrino Factory at J-PARC}

In the final section of my report, I would like to introduce our 
studies on the possibility of the future very long baseline (VLBL) 
neutrino oscillation experiments with a neutrino factory at J-PARC 
and a BAND like 100~kton level detector at a few thousand km away.  
The possibility of a neutrino factory at J-PARC
is being studied in Ref.~\cite{NF_Japan}.
For definiteness, we assume a 100 kton-level segmented
water-\cerenkov calorimeter detector BAND at $L=2,100$km 
away from J-PARC, 
and a neutrino factory at J-PARC which is capable of delivering 
$10^{21}$ $\mu^+$ or $\mu^-$ decays at 10~GeV in one year.  
We will show that the goals of determining all the neutrino model 
parameters, including the neutrino mass hierarchy, 
the sign of $\sin^2\theta_{_{\rm ATM}}-1/2$,
and the degeneracy in the CP phase $\delta_{_{\rm MNS}}$, 
can be reached if such an experiment can be realized.  

The physics prospects of VLBL oscillation experiments with 
a neutrino factory has been studied in the past by assuming that 
the detector can identify charges \cite{NeutrinoFactory}, 
and hence the possibility of charge identification at BAND 
has been investigated \cite{BAND}.  
We have tried to show in the report \cite{nf2b} that even if 
the detector is charge blind, the VLBL experiment will achieve 
all the physics goals if it is capable of distinguishing 
the $\mu$-like and $e$-like events with high confidence level, 
and if the detector is capable of determining the event energy 
(the incoming neutrino energy minus the outgoing neutrino energies) 
at the 10~\% level.  
In the following, we assume that the proposed detector BAND will 
have such capabilities.  

In the neutrino factory, neutrinos are produced 
from the decay of high energy muons, 
$\mu^+ \to \ov\nu_\mu \nu_e e^+$ or $\mu^- \to \nu_\mu \ov\nu_e e^- $.
Not only the shape and the ratios of the neutrino fluxes but 
also their overall normalization will be known accurately in 
the neutrino factory experiments.  
Assuming the relativistic muons, the $\ov \nu_\mu$ and $\nu_e$ 
($\nu_\mu$ and $\ov\nu_e$ ) fluxes from $\mu^+ (\mu^-)$ beam
are expressed as 
\bseq
\label{nufact-flux}
\bea
\Phi_{\ov \nu_\mu (\nu_\mu)} &=&
\frac{\gamma^2 n_\mu}{\pi L^2}2y^2
[(3-2y)\mp P_\mu(1-2y)] \,,\qquad
\eqlab{flux_mu} 
\\
\Phi_{\nu_e (\ov \nu_e)}
&=&\frac{\gamma^2n_\mu}{\pi L^2} 12y^2
[(1-y) \mp P_\mu(1-y)] \,, \quad\;
\eqlab{flux_e} 
\eea
\eseq
where $\gamma=E_\nu/m_\mu$, $y=E_\nu/E_\mu$, and 
$P_\mu$ is the average muon polarization, and
$n_\mu$ is the number of the decaying muons for which we assume 
$10^{21}$ per year in the following.

The $e$-like signal from $\mu^+$ beam, $N_e(\mu^+)$,
is given by the sum of
$e^+$ from $\ov\nu_\mu \to \ov\nu_e$ appearance and 
$e^-$ from the $\nu_e \to \nu_e $ survival mode,
whereas the $\mu$-like signal, $N_{\mu}(\mu^+)$,  
is the sum of 
$\mu^+$ from $\ov\nu_\mu \to \ov\nu_\mu$ and 
$\mu^-$ from $\nu_e \to \nu_\mu $; 
\bseq
\label{flux-mu+}
\bea
N_e(\mu^+) \;&:&\; (\nu_e \to\nu_e)\,+\,(\ov\nu_\mu \to \ov\nu_e)\,,
\\
N_\mu(\mu^+) \;&:&\; (\ov\nu_\mu \to \ov\nu_\mu)\,+\,(\nu_e \to\nu_\mu)\,.  
\eea
\eseq
The signals from the $\mu^-$ beam, $N_e(\mu^-)$ and  $N_{\mu}(\mu^-)$,
are obtained in the same way;
\bseq
\label{flux-mu-}
\bea
N_e(\mu^-) \;&:&\; (\ov\nu_e \to\ov\nu_e)\,+\,(\nu_\mu \to \nu_e)\,,
\\
N_\mu(\mu^-) \;&:&\; (\nu_\mu \to \nu_\mu)\,+\,(\ov\nu_e \to\ov\nu_\mu)\,.
\eea
\eseq
These signals in the {\it i}-th energy bin, 
\bea
\frac{i+1}{10}E_{\mu^\pm} < E_{\obs} < \frac{i+2}{10}E_{\mu^\pm} \,,
\eea
are then calculated by ignoring the energy resolution errors, 
but correcting for the missing neutrino energies in the events 
where $e^\pm$ and $\mu^\pm$ comes from $\tau^\pm$ production via 
the charged currents.  
In the following analysis we dropped from our analysis low energy 
events with $E_{\obs} < 2$~GeV.  

\begin{figure}[p]
\begin{center}
\includegraphics[angle=-90,width=5cm]{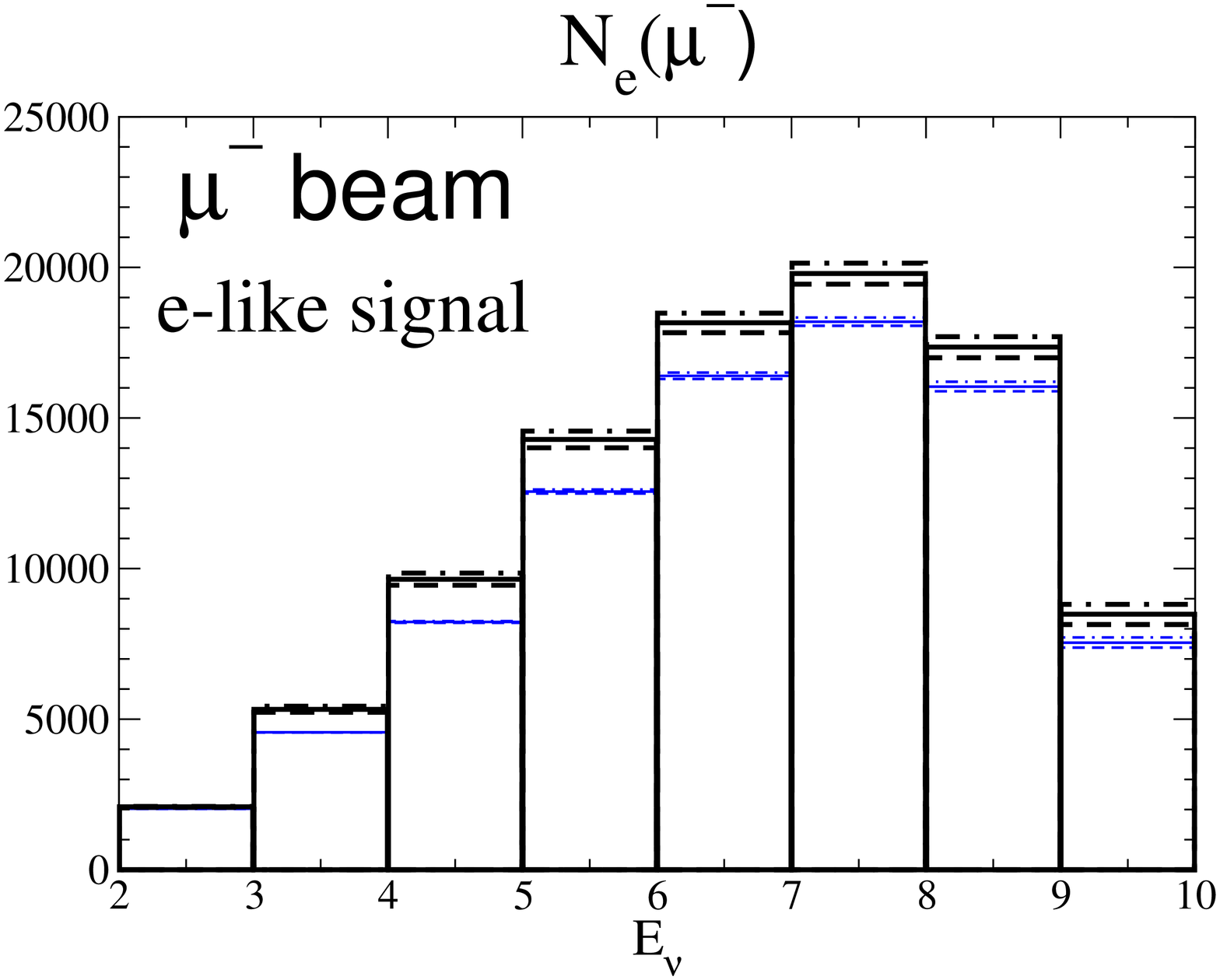}
\includegraphics[angle=-90,width=5cm]{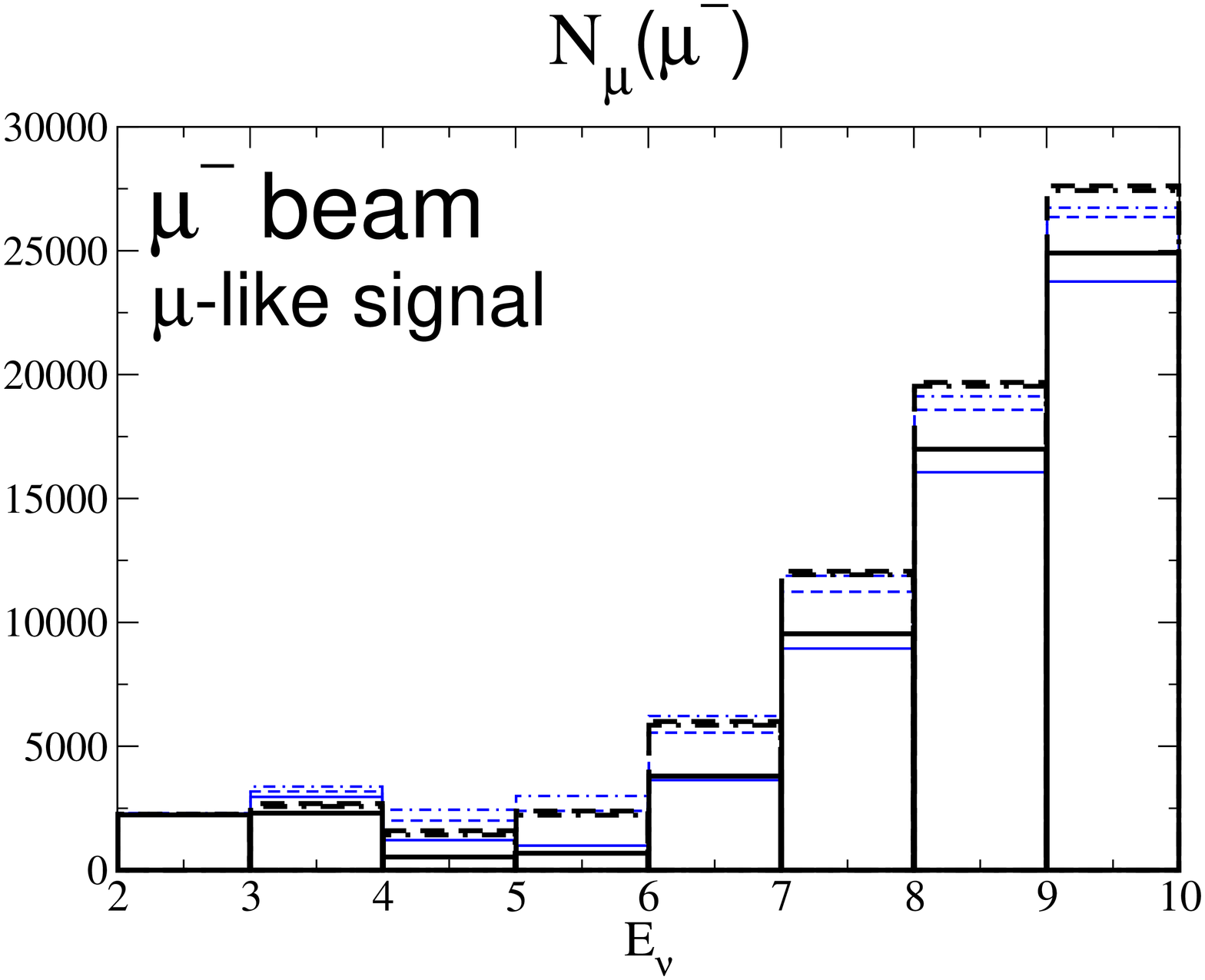}
\includegraphics[angle=-90,width=5cm]{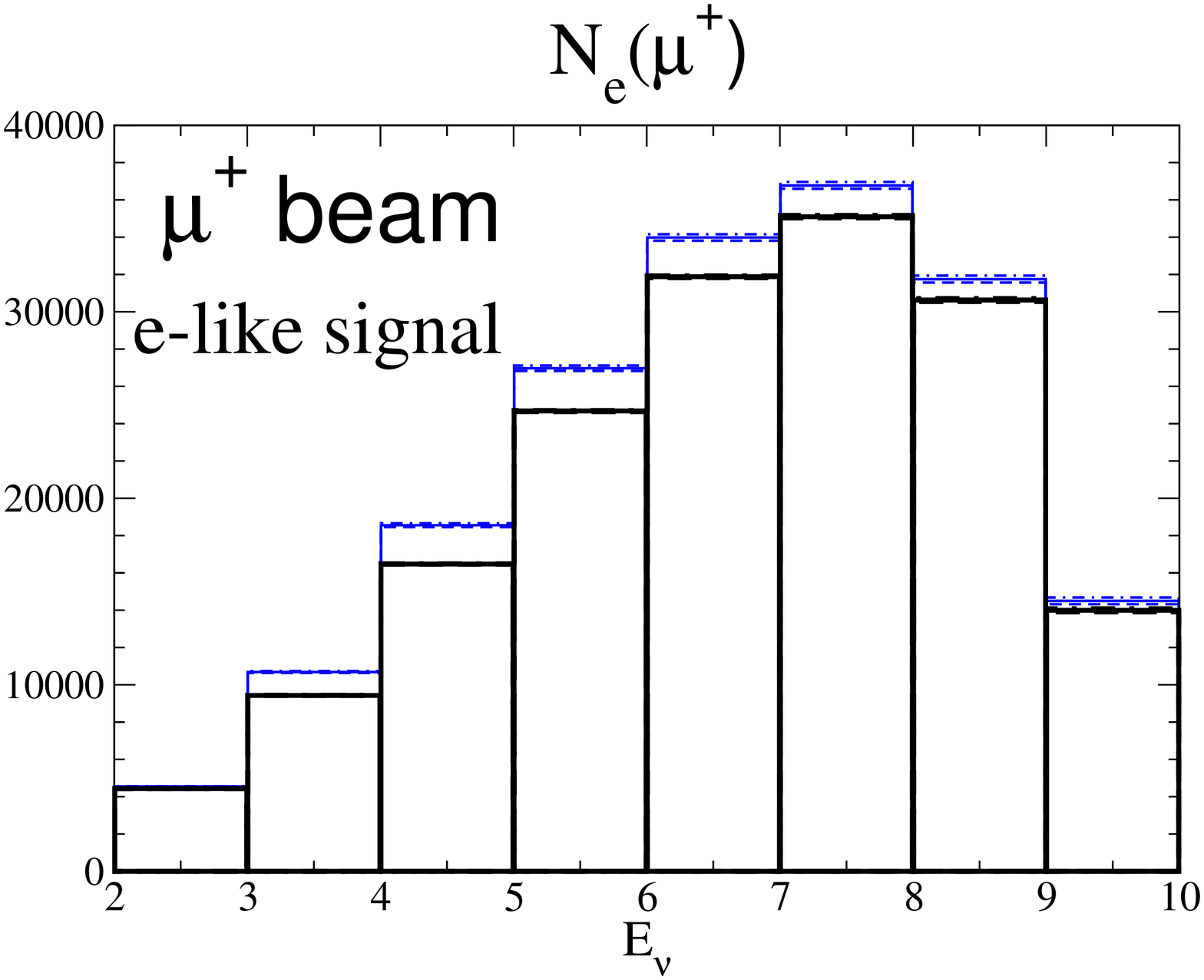}
\includegraphics[angle=-90,width=5cm]{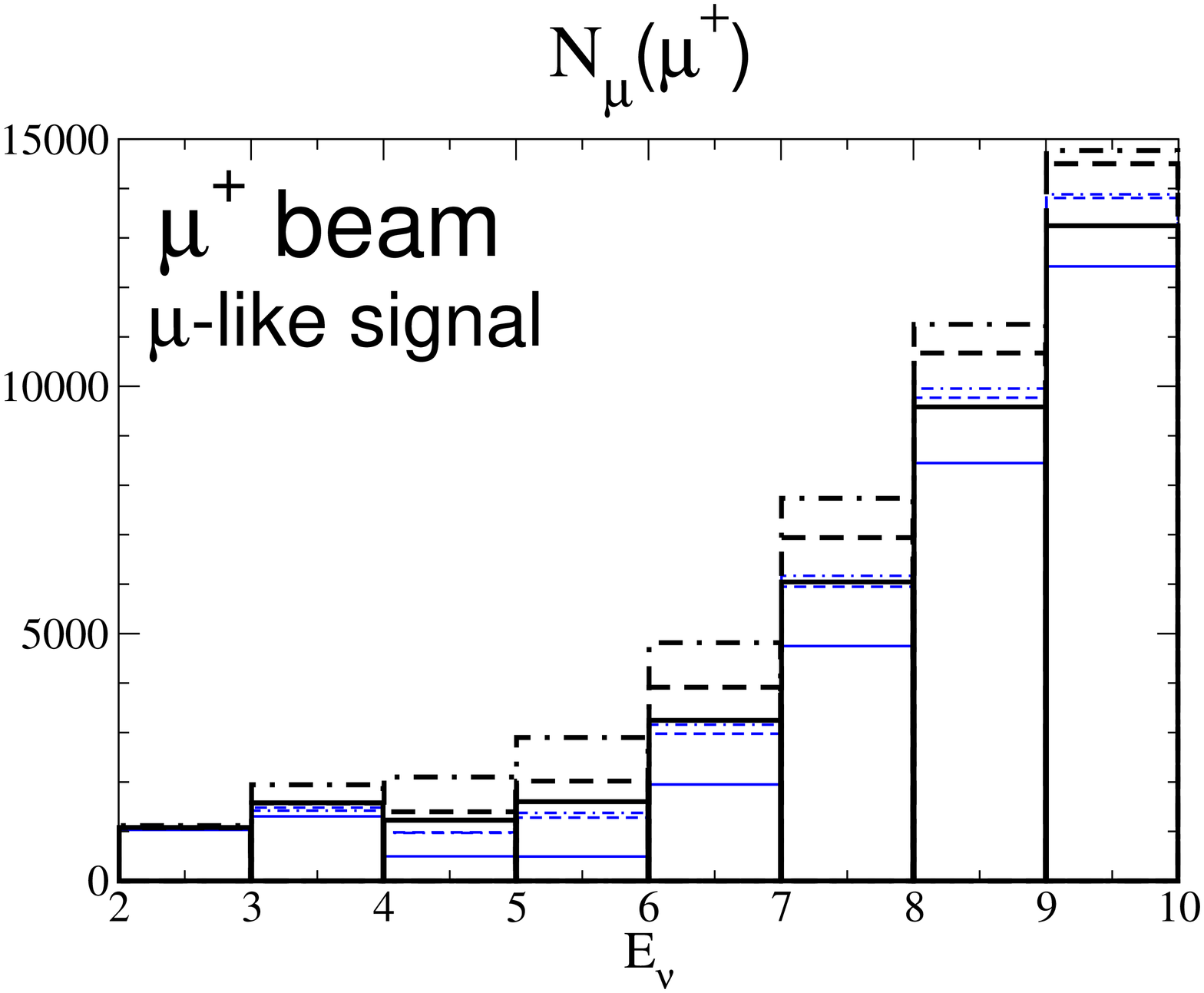}
\caption{
Number of $e$- and $\mu$-like signals for $10^{21}$ decaying 
$\mu^+$ (top two) and $\mu^-$ (bottom two) each at 10 GeV.  
Thick lines are for the normal hierarchy, and thin lines are 
for inverted; 
$\sin^2\theta_{_{\rm ATM}}=$0.5 (solid lines), 0.35 (dashed lines), 
and 0.65 (dot-dashed lines).
}
\label{nf2b-histogram}
\end{center}
\end{figure}

The results are shown in Fig.~\ref{nf2b-histogram} for 
$10^{21}$ decaying $\mu^+$ (top two figures) and 
$\mu^-$ (bottom two figures) each at 10~GeV.
The results for the normal-hierarchy are given in thick lines and 
those of the inverted hierarchy are in thin lines;  
the solid, dashed, and dot-dashed lines are for  
$\sin^2\theta_{_{\rm ATM}}=$0.5, 0.35, and 0.65, respectively. 
The other model parameters are chosen as in eq.~(\ref{t2hk-inputs})  
at $\sin^2 2\thrct=0.06$ and $\dmns = 0^\circ$.  

The contribution from the survival mode dominates the $\mu$-like 
signal but they vanish at around $E_\nu \simeq 5$ GeV for both 
$N_\mu(\mu^+)$ and $N_\mu(\mu^-)$, where most of $\mu$ neutrinos 
oscillate into $\tau$ neutrinos.  
Therefore the $\mu$-like events in the surrounding energy bins are 
sensitive to the $\nu_e \to \nu_\mu$ or $\ov\nu_e \to \ov\nu_\mu$  
transition probability.  

For instance, in the histogram for $N_\mu(\mu^+)$, we can see 
significant dependence on the sign of $\satms{}-1/2$, especially 
around $E_\nu \simeq$ 5 GeV.
The sensitivity is due to the $\nu_e \to \nu_\mu$ appearance mode, 
whose probability is proportional to 
$\sin^2\theta_{_{\rm ATM}}\sin^2\theta_{_{\rm RCT}}$, 
just like the $\nu_\mu \to \nu_e$ oscillation probability 
in eq.~(\ref{pnumu2nue}).  
In order to resolve the degeneracy in the sign of $\satms{}-1/2$, 
however, we need to have an independent measurement of $\sin^2 2\thrct$.  

We observe from Fig.~\ref{nf2b-histogram} that $N_e(\mu^+)$ 
is very insensitive to $\satms{}-1/2$ throughout the energy bins.  
This is especially so for the normal hierarchy, where the 
$\ov\nu_\mu \to \ov\nu_e$ oscillation is strongly suppressed 
due to the matter effects, as explained in the previous section, 
see Fig.~\ref{t2b-1234}.  
Although the suppression is absent for the inverted hierarchy, 
the contributions from the $\ov\nu_\mu \to \ov\nu_e$ transition 
is still suppressed significantly because of the small anti-neutrino 
cross sections on the matter target.  
It is therefore envisaged that the measurement of $N_e(\mu^+)$ 
will give us the desired independent measurement of $\sin^2 2\thrct$ 
because it is dominated by the $\nu_e \to \nu_e$ survival mode, 
especially when the mass hierarchy is normal.  
Therefore, the neutrino factory VLBL experiments with a detector 
capable for distinguishing $e^\pm$ events from $\mu^\pm$ can  
give us a precise measurement of $\sin^2 2\thrct$ just like 
the next generation reactor experiments \cite{chooz2,kaska}.  
In addition, the presence of the enhanced large matter effects 
for the $\nu_\mu \leftrightarrow \nu_e$ and 
$\ov\nu_\mu \leftrightarrow \ov\nu_e$ make the inter-dependences 
of the four observables in Fig.~\ref{nf2b-histogram} very different 
between the normal and the inverted hierarchies.  
We also find that the results depend rather significantly on 
the input $\dmns$ value if $\sin^2 2\thrct$ is not too small.   
Because the energy dependence of the signal can be studied 
in the wide range of $L/E$ at VLBL experiments with a neutrino 
factory, the degeneracy between $\dmns = 0^\circ$ and $180^\circ$ 
can also be resolved \cite{nf2b}.  

We performed a $\chi^2$ analysis similar to the ones we performed 
for the Tokai-to-HK \cite{t2hk} and the Tokai-to-Beijing \cite{t2b} 
projects, in order to study these questions quantitatively.    
The additional assumption that we make for the detector is that 
it is capable of measuring the event energy calorimetrically, 
with the accuracy which makes the error of our analysis based 
on the 1~GeV bin histogram small.  
%We understand that it will indeed be the case for the proposed 
%segmented water \cerenkov calorimeter detector BAND \cite{BAND}.  
%The following results are obtained for 100 kton$\cdot$year 
%each for $10^{21}$ unpolarized $\mu^+$ and $\mu^-$ decays 
%per a year at $E_\mu=10$ GeV and at $L=2,100$~km.   
The energy threshold of $E_\nu>2$~GeV has been introduced to avoid 
numerical sensitivity to the binning due to rapid oscillation.  
Our analysis is hence performed for 8 bins between $E_\nu=$ 2 GeV 
and 10 GeV.

Because the expected number of events is huge for the above 
experimental set up, a few tenths of sounds of events for each bin, 
as can be seen from Fig.~\ref{nf2b-histogram}, 
we introduced the following systematic uncertainties in 
the analysis: 
\begin{itemize}
\item{
2~\% error each in the detection efficiency of $e$-like and $\mu$-like events 
}
\item{
2~\% error each in the CC cross section of neutrinos and anti-neutrinos
}
\end{itemize}
Here we assumed that the errors in the detection efficiencies are 
independent for $e$-like and $\mu$-like events, and also the 
errors in the $\nu$ and $\ov\nu$ cross sections are independent.  
On the other hand we assume that the errors for $\nu_e$ and $\nu_\mu$ 
cross sections are common (100~\% correlated), as well as for their 
anti-neutrino counterparts.  
Since the experiments will be performed more than a decade in the 
future, it is possible that the systematic errors can be reduced 
even further.  
In ref.~\cite{nf2b}, we also show the results when the 
above systematic errors are set to zero.  

\begin{figure}[t]
\begin{center}
\includegraphics[angle=-90,width=7cm]{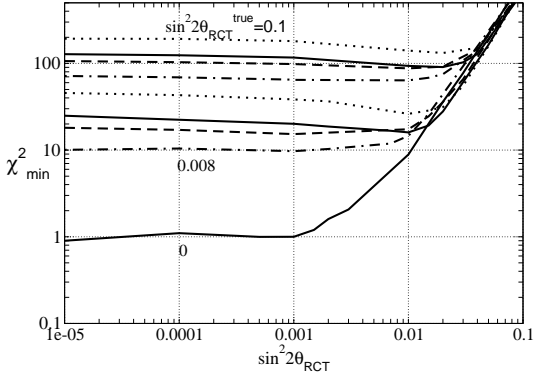}
\caption{%
$\chi^2_{\mbox{\scriptsize \rm min}}$ as a function of the fitting parameter 
$\sin^2 2\theta_{_{\rm RCT}}$ when the input events are generated  
for the normal hierarchy (I) and the analysis is done by assuming 
the inverted hierarchy (III).    
The input data are calculated for $\sin^2\theta_{_{\rm ATM}}^{\true}=$0.5 
at $\sin^22\theta_{_{\rm RCT}}^{\true} =0.1$, 0.008 and 0.  
The four curves are for $\dmns = 0^\circ$ (solid), 
$90^\circ$ (dotted), $180^\circ$ (dash) and $270^\circ$ (dot-dash).  
}
\label{nf2b-hierarchy}
\end{center}
\end{figure}

We show in Fig.~\ref{nf2b-hierarchy} the \chisqmin value 
as a function of $\sin^22\theta_{_{\rm RCT}}$, 
when the `data' are generated by assuming the normal hierarchy, 
while the analysis is done by assuming the inverted hierarchy.  
We show our results for 
$\sin^22\theta_{_{\rm RCT}}^{\true} = 0.1$, 0.008, and 0, 
and for the four CP violating phase angles; $\dmns=0^\circ$, 
$90^\circ$, $180^\circ$ and $270^\circ$.  
The results show that we can distinguish between 
the normal and the inverted mass hierarchy at 3-$\sigma$ level, 
if $\srct{\true} \gsim 0.008$.  
At small $\srct{}$, we find that the contributions of the transition 
modes become negligible, but the $e$-like signals from both 
$\mu^+$ and $\mu^-$ beams are useful in determine the hierarchy.  
This is because the disappearance probability is suppressed for
the $\ov \nu_e \to \ov \nu_e $ transition in the normal hierarchy,
whereas that of $\nu_e \to \nu_e$ is suppressed in the inverted
hierarchy.  
However, the differences of the $e$-like signals between the normal 
and the inverted hierarchy reduces to $\sim 1\%$ when 
$\sin^22\theta_{_{\rm RCT}}=0.004$, and the sensitivity is 
limited by our knowledge on the ratio 
$\sigma_{\cc}^{}(\ov\nu_e)
/\sigma_{\cc}^{}(\nu_e)$.  

\begin{figure}[t]
\begin{center}
\includegraphics[angle=-90,width=7cm]{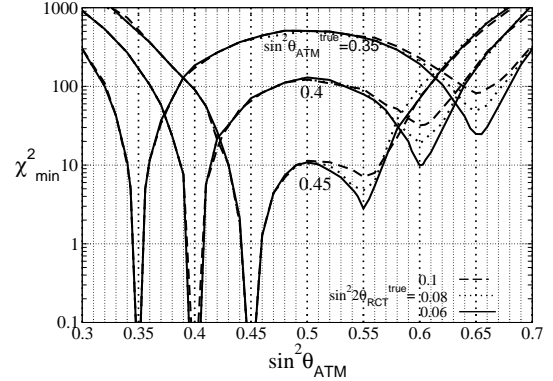}
\caption{%
$\chi^2_{\mbox{\scriptsize \rm min}}$ as a function of the fitting
$\sin^2\theta_{_{\rm ATM}}$.
The input data 
of the events are calculated for 
$\sin^2\theta_{_{\rm ATM}}^{\true}=$0.35, 0.4, and 0.45
with $\sin^22\theta_{_{\rm RCT}}^{\true} =0.06$ 
(solid lines), 
0.08 (dotted lines) and 0.1 (dashed lines) 
and $\delta_{_{\rm MNS}}^{\true}=0^\circ$.
The other input values are the same as in
eq.~(\ref{t2hk-inputs}).
}
\label{nf2b-atm}
\end{center}
\end{figure}

We show in Fig.~\ref{nf2b-atm} the \chisqmin as a function of
$\sin^2\theta_{_{\rm ATM}}$.   
The input `data' are calculated for 
$\sin^2\theta_{_{\rm ATM}}^{\true}=0.35\,$, 0.4\,, and 0.45\,
for three values of $\sin^22\theta_{_{\rm RCT}}^{\true} =0.1$
(dashed lines), 0.08 (dotted lines) and 0.06 (solid lines), 
and for $\delta_{_{\rm MNS}}^{\true}=0^\circ$ with the normal hierarchy.
The values of the other parameters are taken as in eq.~(\ref{t2hk-inputs}).  
The \chisqmin function is found by varying
the fitting parameters within the normal hierarchy.
We see that each \chisqmin has two dips at the $\sin^2 \thatm$ values 
which give the same $\sin^2 2\thatm$.  
The results show that we can resolve the degeneracy at the 3-$\sigma$ 
level when $\sin^2\theta_{_{\rm ATM}}^{\true}=0.35$ 
($\sin^2 2\thatm=0.91$) or 0.4 ($\sin^2 2\thatm=0.96$)  
for all the three $\sin^2 2\theta_{\rct}^{\true}$ values considered 
($\sin^2 2\theta_{\rct}^{\true} \ge 0.06)$, 
but that $\sin^2 \theta_{_{\rm ATM}}^{\true}=0.45$ 
($\sin^2 2\theta_{\atm}^{\true} = 0.99$) 
can be resolved from $\sin^2 \thatm=0.55$ at 3-$\sigma$ 
only if $\sin^2 2\thrct \gsim 0.1$.  

\begin{figure}[p]
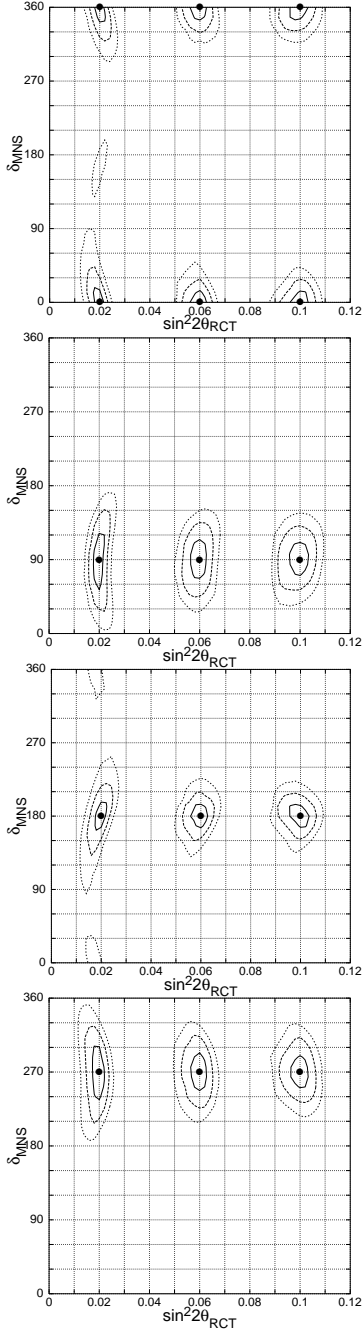

\begin{center}
\includegraphics[angle=-90,width=4.7cm]{nf2b_000.eps}
\includegraphics[angle=-90,width=4.7cm]{nf2b_090.eps}
\includegraphics[angle=-90,width=4.7cm]{nf2b_180.eps}
\includegraphics[angle=-90,width=4.7cm]{nf2b_270.eps}
\caption{%
Allowed regions in the plane of $\sin^22\theta_{_{\rm RCT}}$
and $\delta_{_{\rm MNS}}$ for 
$\sin^22\theta_{_{\rm RCT}}^{\true}=$~0.02, 0.06 and 0.1, for  
$\delta_{_{\rm MNS}}^{\true}=0^{\circ}$, $90^{\circ}$, $180^{\circ}$  
and $270^{\circ}$.  
}
\label{nf2b-contour}
\end{center}
\end{figure}

In Fig.~\ref{nf2b-contour}, we show the allowed regions in the plane of 
$\sin^22\theta_{_{\rm RCT}}$ and $\delta_{_{\rm MNS}}$
for $\sin^22\theta_{_{\rm RCT}}^{\true}=0.02\,, 0.06\,,$ and $0.1$ 
and $\delta_{_{\rm MNS}}^{\true}=0^{\circ}$ (top),
$90^{\circ}$ (second), $180^{\circ}$ (third), 
and $270^{\circ}$ (bottom) in the normal hierarchy.  
In each figure, the input parameter points are shown by solid-circles.
The normal hierarchy is assumed in the fitting.
The regions where \chisqmin $< 1$, 4, and 9 are depicted by
solid, dashed, and dotted boundaries, respectively.

%%%%%%%%%%%%%%%%%%%%%%%%%%%%%%%%%%%%%%%%%%%%%%%%%%%%%%%%%%%%%%%%%%%%%%%%%%%%
%%%%%%%%%%%%%%%%%%%%%%%%%%%%%%%%%%%%%%%%%%%%%%%%%%%%%%%%%%%%%%%%%%%%%%%%%%%%
%%%%%%%%%%%%%%%%%%%%%%%%%%%%%%%%%%%%%%%%%%%%%%%%%%%%%%%%%%%%%%%%%%%%%%%%%%%%

The figures for
$\delta_{_{\rm MNS}}^{\true}=90^\circ$ and 
$\delta_{_{\rm MNS}}^{\true}=270^\circ$ show that 
the CP phase $\delta_{_{\rm MNS}}$ can be constrained locally
around the `true' points for all the input 
$\sin^2 2\thrct$ values.  
We can hence establish CP violation in the leptonic sector, 
and discriminate the maximal CP violation, 
$\delta_{_{\rm MNS}}^{\true}=90^\circ$ or $270^\circ$, 
from the CP conserving cases of $\delta_{_{\rm MNS}}=0^\circ$ or 
$180^\circ$ at the 3-$\sigma$ level if 
$\sin^22\theta_{_{\rm RCT}}^{\true}\gsim 0.02$.
The discrimination between $\dmns = 0^\circ$ and $180^\circ$ 
is also possible at the same level, but for the small 
$4<$ \chisqmin $<9$ island at 
$\sin^22\theta_{_{\rm RCT}}^{\true} = 0.02$.

The above results for CP violation are comparable to those of 
Tokai-to-HK studies shown in Fig.~\ref{fig:t2hk}. 
The sensitivity in this analysis, with the neutrino factory and 
a charge-blind detector, is based essentially on the T violating 
difference between $P_{\nu_e \to \nu_\mu}$ and $P_{\nu_\mu \to \nu_e}$. 
Because the matter effects enhance both of these transitions whereas
suppress strongly the transitions $\ov \nu_\mu \to \ov \nu_e$ and
$\ov \nu_e\to \ov \nu_\mu$, 
$N_\mu(\mu^+)$ is sensitive to $P_{\nu_e \to \nu_\mu}$ and
$N_e(\mu^-)$ is sensitive to $P_{\nu_\mu \to \nu_e}$. 
Therefore, the above results are obtained only when the neutrino 
mass hierarchy is normal.  
A significantly less accurate results are expected if the mass 
hierarchy turns out to be inverted, because the enhanced 
transition modes appear for the anti-neutrinos whose 
cross sections on the matter is a factor of 3 smaller than 
the neutrino cross sections.  

%%%%%%%%%%%%%%%%%%%%%%%%%%%%%%%%%%%%%%%%%%%%%%%%%%%%%%%%%%%%%%%%%%%%%%%%%%%%
%%%%%%%%%%%%%%%%%%%%%%%%%%%%%%%%%%%%%%%%%%%%%%%%%%%%%%%%%%%%%%%%%%%%%%%%%%%%
%%%%%%%%%%%%%%%%%%%%%%%%%%%%%%%%%%%%%%%%%%%%%%%%%%%%%%%%%%%%%%%%%%%%%%%%%%%%
Summing up, the following results are expected for a VLBL experiment 
with a neutrino factory which delivers $10^{21}$ decaying $\mu^+$ and 
$\mu^-$ at 10 GeV and a 100~kton detector which is placed 2,100~km away 
and is capable of measuring the event energy and distinguishing $e^\pm$ 
from $\mu^\pm$, but not their charges:  
The neutrino mass hierarchy can be determined if 
$\sin^2 2\thrct \gsim 0.008$, 
the degeneracy in $\sin^2 \thatm$ can be resolved  
for $\sin^22\theta_{_{\rm ATM}}=0.96$ 
if $\sin^2 2\thrct \gsim 0.06$, 
and the CP-violating phase $\delta_{_{\rm MNS}}$ can be uniquely 
constrained for $\sin^2 2\thrct \gsim 0.02$ if its true value is 
around $90^\circ$ or $270^\circ$, while 
it can be constrained for $\sin^22\theta_{_{\rm RCT}}\gsim 0.04$
if its true value is around $0^\circ$ or $180^\circ$, 
all at the 3-$\sigma$ level. 
%%%%%%%%%%%%%%%%%%%%%%%%%%%%%%%%%%%%%%%%%%%%%%%%%%%%%%%%%%%%%%%%%%%%%%%%%%%%
%%%%%%%%%%%%%%%%%%%%%%%%%%%%%%%%%%%%%%%%%%%%%%%%%%%%%%%%%%%%%%%%%%%%%%%%%%%%
%%%%%%%%%%%%%%%%%%%%%%%%%%%%%%%%%%%%%%%%%%%%%%%%%%%%%%%%%%%%%%%%%%%%%%%%%%%%

{\it Acknowledgments}\\
Let me first thank my collaborators, Mayumi Aoki, Naotoshi Okamura, 
and Ken-ichi Senda.  All the results presented in this report are 
based on the works done or being done with them. 
I also learned a lot from stimulating discussions with our 
colleagues, Y.~Hayato, T.~Kobayashi, T.~Nakaya and K.~Nishikawa.

\end{document}